\documentclass[prd,superscriptaddress,amsfonts,amssymb,amsmath,showpacs]{revtex4-2}
\usepackage{bm}
\usepackage{amsfonts}
\usepackage{latexsym}
\usepackage[utf8]{inputenc}
\usepackage{graphicx}
\usepackage{amsmath}
\usepackage{palatino}
\usepackage{mathpazo}
\usepackage{textcomp}
\usepackage{float}
\usepackage{booktabs}
\usepackage{dcolumn}
\usepackage{ragged2e}
\usepackage{hyperref}
\hypersetup{colorlinks,citecolor=blue}
\hypersetup{colorlinks=true,linkcolor=blue,filecolor=blue,    urlcolor=purple}
\usepackage{amsmath}
\usepackage{xcolor}
\usepackage{orcidlink}
\usepackage[caption=false]{subfig}
\usepackage{commath}
\def\jnl@style{\it}
\def\aaref@jnl#1{{\jnl@style#1}}

\def\aaref@jnl#1{{\jnl@style#1}}

\def\aj{\aaref@jnl{AJ}}                   
\def\apj{\aaref@jnl{ApJ}}                 
\def\apjl{\aaref@jnl{ApJ}}                
\def\apjs{\aaref@jnl{ApJS}}               
\def\apss{\aaref@jnl{Ap\&SS}}             
\def\aap{\aaref@jnl{A\&A}}                
\def\aapr{\aaref@jnl{A\&A~Rev.}}          
\def\aaps{\aaref@jnl{A\&AS}}              
\def\mnras{\aaref@jnl{Mon.~Not.~Roy.~Astron.~Soc.}}             
\def\prd{\aaref@jnl{Phys.~Rev.~D}}        
\def\plb{\aaref@jnl{Phys.~Lett.~B}}        
\def\prc{\aaref@jnl{Phys.~Rev.~C}}  
\def\prl{\aaref@jnl{Phys.~Rev.~Lett.}}    
\def\qjras{\aaref@jnl{QJRAS}}             
\def\skytel{\aaref@jnl{S\&T}}             
\def\ssr{\aaref@jnl{Space~Sci.~Rev.}}     
\def\zap{\aaref@jnl{ZAp}}                 
\def\nat{\aaref@jnl{Nature}}              
\def\aplett{\aaref@jnl{Astrophys.~Lett.}} 
\def\apspr{\aaref@jnl{Astrophys.~Space~Phys.~Res.}} 
\def\physrep{\aaref@jnl{Phys.~Rep.}}      
\def\physscr{\aaref@jnl{Phys.~Scr}}       
\def\commat{\aaref@jnl{Comm.~Math.~Phys.}}              
\def\science{\aaref@jnl{Science}}               
\def\cqg{\aaref@jnl{Classical Quant.~Grav.}}            
\def\jpcs{\aaref@jnl{JPCS}}                                     
\def\ijmpd{\aaref@jnl{Int.~J.~Mod.~Phys.~D}}                    
\def\grg{\aaref@jnl{Gen.~Relat.~Gravit.}}               
\def\rpp{\aaref@jnl{Rep.~Prog.~Phys.}}          
\def\npa{\aaref@jnl{Nucl.~Phys.~A}}        
\def\lrr{\aaref@jnl{Living Rev.~Rel.}}                   
\def\jcap{\aaref@jnl{J.~Cosmology Astropart.~Phys.}}    
\def\rmp{\aaref@jnl{Rev.~Mod.~Phys.}}   
\def\epjc{\aaref@jnl{Eur.~Phys.~J.~C}}

\allowdisplaybreaks[1]
\renewcommand{\arraystretch}{1.1}
\begin{document}

\color{black}       

\title{Energy–Momentum–Squared Gravity:\\
Charged Quark Star Solutions with Unified Interacting Matter }

\author{Ayan~Banerjee} 
\email{ayanbanerjeemath@gmail.com}
\affiliation{Astrophysics and Cosmology Research Unit, School of Mathematics, Statistics and Computer Science, University of KwaZulu--Natal, Private Bag X54001, Durban 4000, South Africa}

\author{Bobur~Turimov\orcidlink{0000-0003-1502-2053}} 
\email{bturimov@astrin.uz}
\affiliation{Central Asian University, Milliy Bog Str. 264, Tashkent, 111221, Uzbekistan}\affiliation{University of Tashkent for Applied Sciences, Str. Gavhar 1, Tashkent, 100149, Uzbekistan}

\author{Javlon~Rayimbaev}
\email{javlonrayimbaev6@gmail.com}
\affiliation{Institute of Theoretical Physics, National University of Uzbekistan, Tashkent 100174, Uzbekistan}
\affiliation{Tashkent International University of Education, Imom Bukhoriy 6, Tashkent 100207, Uzbekistan}
\affiliation{School of Physics, Harbin Institute of Technology, Harbin 150001, China}

\author{Faisal Javed}
\email{faisaljaved.math@gmail.com }\affiliation{College of Transportation, Tongji University, Shanghai 201804, People's Republic of China}
\affiliation{Research Center of Astrophysics and Cosmology, Khazar University, Baku, AZ1096, 41 Mehseti Street, Azerbaijan}


\author{Gulnoza Palvanova} \email{gulnozaps28@gmail.com} \affiliation{National University of Uzbekistan, Tashkent 100174, Uzbekistan}

\author{Sulton~Usanov}
\email{usanovsulton@gmail.com}
\affiliation{Kimyo International University in Tashkent, Usman Nasyr Str. 156, Tashkent 100121, Uzbekistan}


\date{\today}

\begin{abstract}
We investigate the equilibrium structure of electrically charged quark stars within the framework of energy-momentum-squared gravity (EMSG), adopting a unified interacting quark-matter equation of state that incorporates perturbative QCD corrections, color superconductivity, and finite-strange-quark-mass effects. By solving the modified Tolman-Oppenheimer-Volkoff equations over a broad range of EMSG coupling strengths and interaction parameters, we show that the inclusion of electric charge substantially increases both the maximum mass and radius compared with neutral configurations. For strong negative couplings and pronounced interaction strengths, the resulting stellar models reach masses close to $3\,M_{\odot}$ with compactness ratios of order $0.34$, placing them within the mass range inferred for the secondary component of GW190814, whose nature remains observationally undetermined. The presence of charge reduces the central density at maximum mass by approximately $10$--$15\%$ while maintaining full compliance with causality requirements. Furthermore, comparisons with observational constraints from PSR J1614$-$2230, J0348$+$0432, J0740$+$6620, J0952$-$0607, and recent NICER radius measurements indicate that charged EMSG configurations remain consistent with current astrophysical bounds over a wide parameter space. These results demonstrate that a moderate electric charge, when combined with nonlinear matter-geometry coupling, can support quark stars that are significantly more massive than those predicted in general relativity, underscoring the potential observational relevance of charged compact objects in modified gravity and multimessenger astrophysics.
\end{abstract}

\maketitle

 \section{Introduction}

Compact astrophysical objects—white dwarfs, neutron stars, and black holes—occupy a central place in modern astrophysics as natural laboratories for ultra-dense matter and strong-gravity phenomena. Their observable properties tell us how matter behaves at supranuclear densities and, at the same time, provide tests of gravitational physics in regimes that cannot be reproduced in the laboratory. Observations across the electromagnetic spectrum and via gravitational waves have therefore made the study of compact remnants a key interface among nuclear physics, quantum chromodynamics (QCD), and relativistic gravity.

Within this broader landscape, neutron stars mark the transition from degenerate electron to degenerate baryon matter. Yet, theoretical considerations strongly suggest that, at sufficiently high densities, hadrons may dissolve into deconfined quarks. This possibility motivates the notion of \emph{quark stars} or \emph{strange stars}, compact configurations in which up, down, and strange quarks provide the dominant pressure support. Early work by Bodmer on ``collapsed nuclei'' and by Witten on the cosmic separation of phases framed strange quark matter as a possible absolute ground state of strongly interacting matter, laying the foundation for modern quark-star scenarios~\cite{PhysRevD.4.1601,Witten:1984rs,Alcock:1986hz}. Recent multimessenger constraints, together with QCD-informed modelling of dense matter, have even provided evidence that massive neutron stars may harbour quark-matter cores, strengthening the case for deconfinement at the highest central densities~\cite{Annala:2019puf,Weissenborn:2011qu,Alford:2004pf}. Detailed studies of medium effects in strange quark matter and the role of interactions in strange-star configurations further illustrate how microphysical input can radically alter compact-star phenomenology~\cite{Schertler:1996tq,Holdom:2017gdc}.

To capture these effects in a controlled but flexible manner, several interaction-based frameworks have been developed to describe quark matter at high density. A particularly versatile approach is the unified interacting quark-matter model proposed by Zhang and Mann, which incorporates perturbative QCD corrections, colour superconductivity, and finite-strange-quark-mass contributions within a thermodynamically consistent parametrization that smoothly interpolates between different quark phases~\cite{Zhang:2020jmb}. This unified interacting matter (UIM) framework has been applied to a variety of settings, including anisotropic interacting quark stars, universal relations, and confrontations with recent astrophysical observations~\cite{Pretel:2023nlr,Pretel:2024pem,Tangphati:2024ycu,Errehymy:2024tqr}. Extensions to modified gravity, such as charged quark stars and extreme compact objects in regularized four-dimensional Einstein–Gauss–Bonnet gravity, further demonstrate the utility of interaction-based equations of state when exploring the whole space of relativistic compact-star models~\cite{Gammon:2024gij}. In the present work, the UIM description provides the microphysical foundation for constructing charged quark-star configurations in energy–momentum–squared gravity.

In parallel with these microphysical developments, a broad spectrum
of modified gravity theories has been proposed to address potential
departures from General Relativity (GR) in regimes where curvature
and matter densities become extreme, including $f(R)$
gravity~\cite{Sotiriou:2008rp,DeFelice:2010aj}, $f(R,T)$
gravity~\cite{Harko:2011kv,Myrzakulov:2012qp}, and symmetric
teleparallel generalizations~\cite{Xu:2019sbp,Koussour:2025aop,Myrzakulov:2024kuu,Myrzakulov:2024twj}.
Within this broader landscape, Energy--Momentum--Squared Gravity
(EMSG) provides a particularly economical and well-motivated
framework in which quadratic contractions of the energy--momentum
tensor, $T_{\mu\nu}T^{\mu\nu}$, supplement the Einstein--Hilbert
action and introduce density- and pressure-dependent corrections
that become most relevant in the ultra-dense interiors of neutron
and quark stars~\cite{Roshan:2016mbt,Katirci:2013okf}. In addition
to modifying the effective gravitational coupling, EMSG generally
relaxes the covariant conservation of the matter energy--momentum
tensor, giving rise to an extra force associated with nonminimal
matter--geometry interaction that may leave observable signatures
across astrophysical and cosmological
settings~\cite{Roshan:2016mbt,Board:2017ign,Bahamonde:2019urw}.

Of particular relevance to the present work are applications of EMSG to stellar structure. Neutron-star analyses have placed direct bounds on the EMSG coupling from realistic hadronic equations of state, demonstrating that the quadratic matter term can appreciably shift mass--radius relations and maximum masses relative to GR while remaining consistent with observed pulsar properties~\cite{Akarsu:2018zxl,Nari:2018aqs}. Subsequent studies have extended these investigations to isotropic and anisotropic quark stars, colour-flavour-locked configurations, and interacting quark matter, showing that EMSG can support heavier and more compact stars than in standard gravity~\cite{Singh:2020bdv,Tangphati:2022acb,Sharif:2022mdp,ZeeshanGul:2024vgu,Tangphati:2023dgi,Dayanandan:2025vyw}. Charged quark stars in EMSG have also been explored, indicating that the combined effect of electric fields and nonlinear matter–geometry coupling further enhances the attainable masses and modifies stability properties in a way that may be relevant for the heaviest known compact objects~\cite{Pretel:2023avv}. These developments establish EMSG as a well-motivated framework for revisiting the equilibrium and stability of quark stars and provide the theoretical backdrop for the charged, interacting quark-star models considered in this work. More recent studies have further sharpened the EMSG compact-object picture, including neutron-star structure and curvature diagnostics across hadronic and hadron--quark equations of state~\cite{Ghosh:2025yxh}, universal relations linking tidal deformability, compactness, and oscillation frequencies for (proto-)neutron stars~\cite{Ghosh:2026csk}, anisotropic stellar interiors constructed through gravitational decoupling~\cite{Sharif:2024jjo}, and charged solutions in EMSG confronted with Event Horizon Telescope observations~\cite{Aliyan:2024xwl}.

The possibility that compact stars may carry a net electric charge introduces additional physical ingredients that become particularly relevant in ultra-dense regimes. In such systems, charge separation processes, enhanced electromagnetic pressure, and modified collapse dynamics can all influence the balance between gravity and internal stresses. The general-relativistic framework for hydrostatic equilibrium with charge was established in early analyses of relativistic charged fluid spheres and collapsing configurations, which showed that Coulomb repulsion and electromagnetic self-energies can delay or even prevent gravitational collapse in highly compact objects~\cite{Bekenstein:1971ej,Ghezzi:2005iy,Ray:2003gt}. Subsequent studies have derived mass--radius and compactness bounds for a variety of charged matter models, including incompressible and polytropic equations of state, and demonstrated that a modest net charge can relax classical limits such as the Buchdahl bound and give rise to quasiblack-hole-like configurations that remain regular in their interiors~\cite{Mak:2001ie,Boehmer:2007gq,Arbanil:2013pua,Arbanil:2014usa,Lemos:2014lza,Morales:2018nmq,Kumar:2018hgm}. When the stellar interior is composed of deconfined quark matter, models of electrically charged strange quark stars indicate that charge separation in the outer layers and the associated electromagnetic pressure can significantly modify equilibrium conditions, leading to higher maximum masses, altered mass--radius relations, and distinct oscillatory behaviour compared with neutral quark-star sequences~\cite{Negreiros:2009fd,Arbanil:2015uoa}. These investigations of charged neutron and quark stars therefore suggest that electric charge is a natural additional degree of freedom in the ultra-dense regime and that its inclusion may be essential for a complete description of the heaviest compact objects, especially when considered together with interaction-based quark-matter equations of state and nonlinear gravity effects such as those predicted by EMSG.

A comprehensive framework that combines charged stellar configurations, quark-matter-based compact stars, and modified-gravity theories such as EMSG provides a particularly rich setting for probing strong-field physics. On the microphysical side, unified interacting descriptions of quark matter capture the roles of QCD interactions, colour superconductivity, and the strange-quark mass in determining the stiffness of the equation of state. On the gravitational side, EMSG introduces nonlinear couplings to the energy–momentum tensor that become increasingly important at high densities. At the same time, electric charge supplies an additional source of pressure support and modifies classical compactness bounds~\cite{Akarsu:2018zxl,Tangphati:2022acb,Negreiros:2009fd,Arbanil:2015uoa,Pretel:2023avv}. The interplay between these ingredients can yield novel equilibrium sequences that reach the mass and compactness ranges suggested by the most massive observed compact stars, potentially addressing tensions between data and purely hadronic GR models. This unified perspective serves as the central motivation of the
present work, which investigates charged quark stars composed of
unified interacting quark matter within the EMSG framework and
confronts the resulting configurations with current astrophysical
constraints.  While previous studies have examined
neutral quark stars in EMSG~\cite{Singh:2020bdv,Tangphati:2022acb,Dayanandan:2025vyw}
or charged quark stars with simpler equations of
state~\cite{Pretel:2023avv}, the present work is the first to
combine the unified interacting quark-matter equation of state
with electric charge and EMSG nonlinear gravity, yielding stellar
configurations that extend significantly beyond the predictions of
general relativity and existing modified gravity models in both
maximum mass and compactness.

This work is organized as follows. In Sec.~\ref{sec:fieldequation} we briefly outline the EMSG framework and derive the modified field equations for the Maxwell–EMSG system, together with the corresponding stellar structure equations for static, spherically symmetric configurations. In Sec.~\ref{sec:EoS} we introduce the unified interacting quark-matter equation of state and the charge-density prescription employed to model electrically charged quark stars. Sec.~\ref{sec:numerical} presents the numerical solutions for neutral and charged configurations, discussing their global properties in terms of the EMSG coupling, interaction parameter, and charge fraction. In Sec.~\ref{sec:stability} we examine the static stability criterion, the adiabatic index, and the sound-speed profiles in order to assess dynamical stability and causality. Finally, Sec.~\ref{sec:conclusion} summarizes the main results and outlines prospects for future work.


\section{Framework of Energy–momentum Squared Gravity}\label{sec:fieldequation} 

\subsection{Field equations}

To investigate the equilibrium and structure of charged compact stars within the framework of EMSG, we first outline the gravitational theory and incorporate electromagnetic contributions. The Einstein-Hilbert action is generalized to include an additional term quadratic in the energy-momentum tensor:
\begin{equation}\label{action}
    S = \int d^4x\sqrt{-g} \left[ \frac{R}{16\pi} + \alpha T_{\mu\nu}T^{\mu\nu} + \mathcal{L}_m + \mathcal{L}_e \right] ,
\end{equation}
where $g$ denotes the determinant of the metric tensor, $R$ represents the Ricci scalar curvature, and $T_{\mu\nu}$ is the energy-momentum tensor associated with the matter Lagrangian density $\mathcal{L}_m$. The electromagnetic sector enters through its Lagrangian $\mathcal{L}_e$, while the parameter $\alpha$ quantifies the deviation from general relativity and measures the coupling strength of the EMSG correction. In the present work, the quadratic EMSG invariant $T_{\mu\nu}T^{\mu\nu}$ is constructed from the material perfect-fluid energy--momentum tensor alone. The electromagnetic field is kept out of the squared term and enters only through the Maxwell Lagrangian $\mathcal{L}_e$ and its stress tensor $\mathcal{E}_{\mu\nu}$. In particular, $T_{\mu\nu}T^{\mu\nu}$ contains no $\mathcal{E}_{\mu\nu}\mathcal{E}^{\mu\nu}$ contribution and no mixed matter--electromagnetic contractions.

Variation of this action with respect to the metric yields the modified field equations:
\begin{equation}\label{FieldEq}
    G_{\mu\nu} = 8\pi \left( T_{\mu\nu} + \mathcal{E}_{\mu\nu} \right) + 8\pi\alpha \left( g_{\mu\nu} T_{\sigma\rho}T^{\sigma\rho} - 2\Theta_{\mu\nu} \right) ,
\end{equation}
where $G_{\mu\nu}$ denotes the Einstein tensor, $\mathcal{E}_{\mu\nu}$ represents the electromagnetic energy-momentum tensor, and $\Theta_{\mu\nu}$ is a new tensor arising from the quadratic matter coupling:
\begin{equation}\label{ThetaEq}
    \Theta_{\mu\nu} \equiv T^{\sigma\rho}\frac{\delta T_{\sigma\rho}}{\delta g^{\mu\nu}} + T_{\sigma\rho}\frac{\delta T^{\sigma\rho}}{\delta g^{\mu\nu}} = 2T_\mu^\sigma T_{\nu\sigma} - 2\mathcal{L}_m\left[ T_{\mu\nu} - \frac{1}{2}g_{\mu\nu}T \right] - TT_{\mu\nu} - 4T^{\sigma\rho} \frac{\partial^2 \mathcal{L}_m}{\partial g^{\mu\nu} \partial g^{\sigma\rho}}, 
\end{equation}
with $T$ being the trace of $T_{\mu\nu}$. The matter energy-momentum tensor is related to the Lagrangian density through:
\begin{equation}
    T_{\mu \nu} = \frac{-2}{\sqrt{-g}} \frac{\delta(\sqrt{-g}\mathcal{L}_m)}{\delta g^{\mu\nu}} = g_{\mu\nu}\mathcal{L}_m - 2\frac{\partial\mathcal{L}_m}{\partial g^{\mu\nu}} .
\end{equation}

For a charged stellar configuration, we adopt the perfect fluid description for matter combined with an electromagnetic contribution, following the treatment by Ray \textit{et al.}~\cite{Ray:2003gt}. The respective energy-momentum tensors take the forms:
\begin{equation}\label{MatterEMT}
    T_{\mu\nu} = (\rho+ p)u_\mu u_\nu + pg_{\mu\nu} ,
\end{equation}
and 
\begin{equation}\label{ElectEMT}
    \mathcal{E}_{\mu\nu} = \frac{1}{4\pi} \left[ F_{\mu\lambda}g^{\alpha\lambda}F_{\nu\alpha} - \frac{1}{4}g_{\mu\nu}F_{\lambda\sigma}F^{\lambda\sigma} \right] ,
\end{equation}
where $\rho$ represents the energy density, $p$ the pressure, $u^\mu$ the fluid four-velocity, and $F_{\mu\nu}$ the electromagnetic field strength tensor. This tensor is constructed from the four-potential $A_\mu$ via $F_{\mu\nu} = \nabla_\mu A_\nu - \nabla_\nu A_\mu$, where $\nabla_\mu$ denotes covariant differentiation. The Maxwell equations governing the electromagnetic field are:
\begin{align}
    &\frac{1}{\sqrt{-g}} \frac{\partial}{\partial x^\mu}\left( \sqrt{-g}F^{\mu\nu} \right) = -4\pi j^\nu ,  \label{MaxwEq1}  \\
    &\nabla_\sigma F_{\mu\nu} + \nabla_\mu F_{\nu\sigma} + \nabla_\nu F_{\sigma\mu} = 0 ,  \label{MaxwEq2}
\end{align}
where $j^\mu = \rho_{\rm ch}u^\mu$ denotes the four-current density and $\rho_{\rm ch}$ the electric charge density.

To derive the modified stellar structure equations, we employ the static spherically symmetric metric:
\begin{equation}\label{MetricEq}
    ds^2 = g_{\mu\nu}dx^\mu dx^\nu = -e^{2\psi}dt^2 + e^{2\lambda}dr^2 + r^2(d\theta^2 + \sin^2\theta d\phi^2) ,
\end{equation}
where the metric functions $\psi$ and $\lambda$ depend solely on the radial coordinate $r$. This geometry yields $\sqrt{-g}= e^{\psi+ \lambda}r^2\sin\theta$ and $u^\mu = e^{-\psi}\delta_0^\mu$. For a purely electric field configuration, the only non-vanishing component of the field strength tensor is $F^{01} = -F^{10}$, which, through the Maxwell equation (\ref{MaxwEq1}), leads to:
\begin{equation}
    F^{01} = \frac{q(r)}{r^2}e^{-\psi - \lambda} , 
\end{equation}
where the radial charge function is given by:
\begin{equation}\label{ChargeEq}
    q(r) = 4\pi\int _0^r \bar{r}^2\rho_{\rm ch}(\bar{r})e^{\lambda(\bar{r})}d\bar{r} .
\end{equation}

Since $\nabla^\mu G_{\mu\nu}= 0$ identically, the covariant divergence of the field equations (\ref{FieldEq}) reveals the non-conservation of the matter energy-momentum tensor in EMSG:
\begin{equation}
    \nabla^\mu T_{\mu \nu} = -\nabla^\mu\mathcal{E}_{\mu\nu} - \alpha g_{\mu\nu} \nabla^\mu (T_{\sigma\rho} T^{\sigma\rho}) + 2\alpha \nabla^\mu \Theta_{\mu\nu} ,
\end{equation}
which, using Eqs.~(\ref{MatterEMT}) and (\ref{ElectEMT}), can be expressed as:
\begin{equation}\label{NonConservEq}
    \nabla^\mu T_{\mu \nu} = - j^\lambda F_{\lambda\nu} - \alpha \nabla_\nu \left(\rho^2+ 3p^2\right) + 2\alpha \nabla^\mu \Theta_{\mu\nu} .
\end{equation}

The explicit form of $\Theta_{\mu\nu}$ in Eq.~(\ref{ThetaEq}) depends on the choice of matter Lagrangian density $\mathcal{L}_m$. Following established approaches in the literature~\cite{Faraoni:2009rk,Bertolami:2008ab}, we adopt $\mathcal{L}_m = p$ for this analysis, which yields $\Theta_{\mu\nu} = -(\rho^2+ 4\rho p + 3p^2)u_\mu u_\nu$.  While $\mathcal{L}_m = -\rho$ is an equally admissible choice in GR, the selection $\mathcal{L}_m = p$ has been shown to be the physically appropriate prescription for perfect fluids in theories with non-minimal matter-geometry coupling~\cite{Faraoni:2009rk,Bertolami:2008ab}, and is consistently adopted throughout the EMSG literature~\cite{Akarsu:2018zxl,Nari:2018aqs,Tangphati:2022acb,Pretel:2023avv}. Substituting this expression into Eqs.~(\ref{FieldEq}) and (\ref{NonConservEq}), the field equations and energy-momentum non-conservation take the forms:
\begin{align}
    &G_{\mu\nu} = 8\pi\rho \left[ \left( 1+ \frac{p}{\rho} \right)u_\mu u_\nu + \frac{p}{\rho}g_{\mu\nu} \right] + 8\pi\mathcal{E}_{\mu\nu} + 8\pi\alpha\rho^2 \left[ \left( 1+ 3\frac{p^2}{\rho^2} \right)g_{\mu\nu} + 2\left( 1+ 4\frac{p}{\rho}+ 3\frac{p^2}{\rho^2} \right)u_\mu u_\nu \right] ,  \label{FieldEqLmp}  \\
    &\nabla^\mu T_{\mu \nu} = - j^\lambda F_{\lambda\nu} - \alpha \partial_\nu \left(\rho^2+ 3p^2\right) - 2\alpha\left( \rho^2+ 4\rho p+ 3p^2 \right)\Gamma_{0\nu}^0 .  \label{NonConsevEqLmp}
\end{align}

Evaluating these equations within the spherically symmetric geometry defined by Eq.~(\ref{MetricEq}), the $00$ and $11$ components yield:
\begin{align}
    &\frac{1}{r^2}\frac{d}{dr}\left( re^{-2\lambda} \right) - \frac{1}{r^2} = -8\pi\left( \rho + \frac{q^2}{8\pi r^4} \right) - 8\pi\alpha\rho^2\left( 1+ 8\frac{p}{\rho}+ 3\frac{p^2}{\rho^2} \right) ,  \label{Eq16} \\
    &e^{-2\lambda}\left( \frac{2}{r}\psi'+ \frac{1}{r^2} \right) - \frac{1}{r^2} = 8\pi \left( p - \frac{q^2}{8\pi r^4} \right) + 8\pi\alpha\rho^2\left( 1+ 3\frac{p^2}{\rho^2} \right) ,  \label{Eq17}
\end{align}
where primes denote radial derivatives. These expressions reduce to those derived in Ref.~\cite{Akarsu:2018zxl} when electric charge is absent. The radial component of the non-conservation equation (\ref{NonConsevEqLmp}) provides the pressure gradient:
\begin{equation}\label{pPrimeEq}
    p' = -\frac{\rho +p}{1+ 6\alpha p}\left[ 1+ 2\alpha\rho\left( 1+ 3\frac{p}{\rho} \right) \right]\psi' + \frac{qq'}{4\pi r^4(1+ 6\alpha p)} - \frac{2\alpha\rho\rho'}{1+ 6\alpha p} . 
\end{equation}

To cast these structural relations in a more transparent form reminiscent of the standard Tolman-Oppenheimer-Volkoff (TOV) formulation, we introduce a gravitational mass function that quantifies the total mass-energy enclosed within radius $r$. Integrating Eq.~(\ref{Eq16}) yields:
\begin{equation}
    e^{-2\lambda} = 1- \frac{2}{r}\left\lbrace 4\pi\int r^2\rho dr + \frac{1}{2}\int \frac{q^2}{r^2}dr + 4\pi\alpha\int r^2\rho^2\left( 1+ 8\frac{p}{\rho} + 3\frac{p^2}{\rho^2} \right)dr \right\rbrace ,
\end{equation}
which can be expressed compactly as:
\begin{equation}\label{ExpLambda}
    e^{-2\lambda} = 1 - \frac{2m}{r} + \frac{q^2}{r^2} ,
\end{equation}
where the mass function is given by:
\begin{equation}\label{MassFunc}
    m = 4\pi\int r^2\rho dr + \int\frac{qq'}{r}dr + 4\pi\alpha\int r^2\rho^2\left( 1+ 8\frac{p}{\rho} + 3\frac{p^2}{\rho^2} \right)dr .
\end{equation}

This decomposition reveals that the effective gravitational mass arises from three distinct contributions: the conventional matter energy density captured in the first integral, the electromagnetic field energy encoded in the charge distribution, and the modifications induced by the quadratic matter coupling in EMSG, appearing in the third term. Setting $\alpha= 0$ recovers the familiar charged perfect fluid mass function in general relativity \cite{Negreiros:2009fd, Arbanil:2013pua}, while the neutral limit ($q=0$) reproduces the uncharged EMSG result of Ref.~\cite{Akarsu:2018zxl}. 

Employing the relation (\ref{ExpLambda}), the radial field equation (\ref{Eq17}) determines the metric potential gradient:
\begin{equation}\label{PsiPrimeEq}
    \psi' = \left[ \frac{m}{r^2} + 4\pi rp - \frac{q^2}{r^3} + 4\pi\alpha r\rho^2\left( 1+ 3\frac{p^2}{\rho^2} \right) \right]e^{2\lambda} . 
\end{equation}

Combining Eqs.~(\ref{ChargeEq}), (\ref{pPrimeEq}), (\ref{MassFunc}), and (\ref{PsiPrimeEq}), we arrive at the complete system of modified Tolman-Oppenheimer-Volkoff equations governing charged stellar configurations in EMSG \cite{Pretel:2023avv}:
\begin{align}
    \frac{dq}{dr} =&\ 4\pi r^2\rho_{\rm ch} \left( 1- \frac{2m}{r}+ \frac{q^2}{r^2} \right)^{-1/2} ,  \label{TOV1Lmp}  \\  
    \frac{dm}{dr} =&\ 4\pi r^2\rho + \frac{qq'}{r} + 4\pi\alpha r^2\rho^2\left( 1+ 8\frac{p}{\rho}+ 3\frac{p^2}{\rho^2} \right),  \label{TOV2Lmp}  \\ 
    \frac{dp}{dr} =& -\frac{\rho+ p}{1+ 6\alpha p}\left[ 1+ 2\alpha\rho\left( 1+ 3\frac{p}{\rho} \right) \right]  \nonumber  \\
    &\times \left[ \frac{m}{r^2} + 4\pi rp- \frac{q^2}{r^3} + 4\pi\alpha r\rho^2\left( 1+ 3\frac{p^2}{\rho^2} \right) \right]\left( 1- \frac{2m}{r}+ \frac{q^2}{r^2} \right)^{-1}  \nonumber  \\
    &+ \frac{qq'}{4\pi r^4(1+ 6\alpha p)} - \frac{2\alpha\rho\rho'}{1+ 6\alpha p} ,  \label{TOV3Lmp}  \\
    \frac{d\psi}{dr} =& -\frac{1}{\rho+ p}\left[ (1+ 6\alpha p)p' + 2\alpha\rho\rho' - \frac{qq'}{4\pi r^4} \right]\left[ 1+ 2\alpha\rho\left( 1+ 3\frac{p}{\rho} \right) \right]^{-1} .  \label{TOV4Lmp}
\end{align}

These differential equations describe hydrostatic equilibrium in the presence of both electromagnetic fields and EMSG modifications. When $\alpha =0$, the standard general relativistic TOV equations are recovered, while the uncharged limit ($\rho_{\rm ch}= 0$, $q= 0$) reproduces the neutral EMSG results of Ref.~\cite{Akarsu:2018zxl}.

It is instructive to identify the physical origin of each term in
this system. In the mass equation~(\ref{TOV2Lmp}), the first term
$4\pi r^2\rho$ represents the standard contribution of quark matter
energy density, the second term $qq'/r$ encodes the electromagnetic
field energy, and the third term $4\pi\alpha r^2\rho^2(1 + 8p/\rho +
3p^2/\rho^2)$ is a purely EMSG correction arising from the quadratic
$T_{\mu\nu}T^{\mu\nu}$ coupling. Similarly, in the pressure
equation~(\ref{TOV3Lmp}), the factor $(1 + 6\alpha p)^{-1}$ in the
denominator and the term $2\alpha\rho\rho'/(1+6\alpha p)$ are
exclusively of EMSG origin, while the term $qq'/(4\pi r^4)$
originates from the electromagnetic sector. The remaining structure ---
the $(\rho+p)\psi'$ gravitational coupling and the standard pressure
gradient --- are inherited directly from general relativity. When
$\alpha \to 0$, all EMSG corrections vanish identically and the
familiar charged perfect fluid TOV equations of general relativity
are recovered~\cite{Negreiros:2009fd,Bekenstein:1971ej}.

Numerical integration of this system requires specification of an equation of state $p= p(\rho)$ relating pressure to energy density, along with a charge distribution profile $\rho_{\rm ch} = \rho_{\rm ch}(\rho)$, thereby closing the system for the four unknown functions. Regularity at the stellar center demands the boundary conditions:
\begin{align}\label{BC1}
    q(0) &= 0,   &   m(0) &= 0 ,   &   \rho(0) &= \rho_c ,
\end{align}
where $\rho_c$ denotes the central energy density. Integration proceeds outward from the origin until pressure vanishes, which defines the stellar surface at radius $r_{\rm sur}$ satisfying $p(r_{\rm sur})= 0$. 

To determine the metric potential $\psi$ throughout the interior, an additional boundary condition is required at the surface. Taking the trace of the field equations (\ref{FieldEqLmp}) yields the Ricci scalar $R= 8\pi(\rho- 3p)[1- 2\alpha(\rho- p)]$, which vanishes in the exterior vacuum region where $\rho= p =0$. This relies on the convention above that the squared term is built from the fluid tensor alone: had the electromagnetic stress been included in $T_{\mu\nu}T^{\mu\nu}$, it would not vanish in the charged exterior and the external geometry would depart from Reissner--Nordstr\"om. Consequently, the external spacetime remains the Reissner-Nordström solution of general relativity, and continuity of the metric across the surface imposes:
\begin{equation}
    \psi(r_{\rm sur}) = \frac{1}{2}\ln\left[ 1 - \frac{2M}{r_{\rm sur}} + \frac{Q^2}{r_{\rm sur}^2} \right] ,
\end{equation}
where $M \equiv m(r_{\rm sur})$ and $Q \equiv q(r_{\rm sur})$ represent the total gravitational mass and electric charge of the configuration, respectively. We emphasize that $M$ is a metric (gravitational) mass parameter, not a separately conserved matter rest mass. The function $m(r)$ in Eq.~(\ref{MassFunc}) is read from the radial metric component (\ref{ExpLambda}) and already incorporates the matter, electromagnetic, and effective EMSG contributions; thus, although the material energy-momentum tensor is not separately conserved in EMSG [Eq.~(\ref{NonConservEq})], the contracted Bianchi identity $\nabla^\mu G_{\mu\nu}=0$ ensures that the total effective source is conserved and $m(r)$ is well defined. Because the exterior is Reissner--Nordstr\"om, $M\equiv m(r_{\rm sur})$ is the mass parameter of the matched asymptotically flat metric, and hence governs the orbital motion of distant test bodies and binary companions, making it the appropriate quantity for comparison with pulsar mass measurements.


 \section{Equation of state and the charge density profile}\label{sec:EoS}

\subsection{Unified Interacting Equation of State}

A realistic description of deconfined quark matter is essential for modelling ultra-dense compact stars, particularly when such systems are studied within modified gravity frameworks. In this work, we employ the unified interacting equation of state (EoS) developed in~\cite{Zhang:2020jmb}, which incorporates perturbative QCD effects, color superconductivity, and a finite strange-quark mass into a single parametric framework. This construction provides a continuous and thermodynamically consistent representation of the 2SC, 2SC+s, and CFL phases, well-suited for analysing charged quark stars in the context of EMSG theory.

The formulation begins with the thermodynamic potential \cite{Zhang:2020jmb,Alford:2004pf,Weissenborn:2011qu},
\begin{equation} 
\Omega = 
-\,\frac{\xi_4}{4\pi^{2}}\,\mu^{4}
+ \frac{\xi_4 (1-a_{4})}{4\pi^{2}}\,\mu^{4}
- \frac{\xi_{2a}\Delta^{2} - \xi_{2b} m_{s}^{2}}{\pi^{2}}\,\mu^{2}
- \frac{\mu_{e}^{4}}{12\pi^{2}}
+ B_{\rm eff},
\label{eq:Omega_general_single}
\end{equation}
where $\mu$ denotes the averaged quark chemical potential, $\Delta$ represents the gap parameter for color superconductivity, $m_s$ is the strange quark mass, and $B_{\rm eff}$ captures the effective bag constant that encodes the nonperturbative QCD vacuum contribution \cite{Holdom:2017gdc}. The phase-dependent parameters $(\xi_4,\xi_{2a},\xi_{2b})$ take the following values:
\begin{align} \label{eq:coeff_single}
(\xi_4,\xi_{2a}, \xi_{2b}) = \left\{ \begin{array} {ll}
\bigg(\big( \left(\frac{1}{3}\right)^{\frac{4}{3}}+ \left(\frac{2}{3}\right)^{\frac{4}{3}}\big)^{-3},1,0\bigg), & \textrm{2SC phase,}\\
(3,1,3/4), & \textrm{2SC+s phase,}\\
(3,3,3/4),&   \textrm{CFL phase,}
\end{array}
\right.
\end{align}
Using the standard thermodynamic relations $p=-\Omega$ and $\rho=\Omega+\mu n_q$, we can derive the unified interacting EoS by introducing the interaction parameter
\begin{equation}
\lambda = \frac{\xi_{2a}\Delta^{2}-\xi_{2b}m_s^{2}}
{\sqrt{\xi_4\,a_4}},
\label{eq:lambda_single}
\end{equation}
where $a_4$ characterizes the strength of perturbative QCD corrections from one-gluon exchange at $\mathcal{O}(\alpha_s^2)$. The resulting relation between pressure and density takes the compact form
\begin{equation}
p = \frac{\rho - 4B_{\rm eff}}{3}
+ \frac{4\lambda^2}{9\pi^2}
\left[
-1 
+ {\rm sgn}(\lambda)
\sqrt{1 + 
\frac{3\pi^{2}(\rho - B_{\rm eff})}{\lambda^{2}}}
\right],
\label{eq:UQM_single_dimful}
\end{equation}
which smoothly interpolates between the non-interacting MIT bag model limit at $\lambda \to 0$ and the strongly interacting regime at $\lambda \to \infty$, where the EoS approaches $p=\rho - 2B_{\rm eff}$. Here the prefactor $1/3$ multiplies only the bag-model contribution $(\rho-4B_{\rm eff})$, while the interaction contribution is a separate additive term, following the parenthesization of Ref.~\cite{Zhang:2020jmb}. For the positive interaction branch used in the present numerical analysis, expanding the square root for large positive $\lambda$ with $x=3\pi^{2}(\rho-B_{\rm eff})/\lambda^{2}$ gives $\sqrt{1+x}=1+x/2-x^{2}/8+\mathcal{O}(x^{3})$. Hence the interaction term tends first to $\tfrac{2}{3}(\rho-B_{\rm eff})$, and the full equation of state becomes
\[
p = \rho - 2B_{\rm eff} - \frac{\pi^{2}}{2\lambda^{2}}\,(\rho-B_{\rm eff})^{2} + \mathcal{O}(\lambda^{-4}),
\]
which yields $p\to\rho-2B_{\rm eff}$ and $dp/d\rho\to1$ in the strongly interacting limit.

For numerical convenience, we adopt the rescaled variables
\begin{equation}
\bar{\rho} = \frac{\rho}{4B_{\rm eff}}, 
\qquad
\bar{p}   = \frac{p}{4B_{\rm eff}},
\qquad
\bar{\lambda} = \frac{\lambda^{2}}{4B_{\rm eff}},
\label{eq:scaling_single}
\end{equation}
yielding the dimensionless representation
\begin{equation}
\bar{p} = 
\frac{\bar{\rho}-1}{3}
+ \frac{4}{9\pi^{2}}\bar{\lambda}
\left[
-1 
+ {\rm sgn}(\lambda)
\sqrt{
1 + \frac{3\pi^{2}}{\bar{\lambda}}
\left(\bar{\rho}-\frac{1}{4}\right)
}
\right].
\label{eq:UQM_single_dimless}
\end{equation}

The parameter $\bar{\lambda}$ effectively governs the stiffness of the EoS. Larger values correspond to stronger interactions between quarks, enhanced color superconductivity, or reduced strange quark mass. When we couple this EoS to the modified Tolman-Oppenheimer-Volkoff equations in EMSG, we obtain the mass-radius sequences examined later in this work. Notably, the unified parametrization avoids inconsistencies arising from independent tuning of microscopic parameters, ensuring that the resulting stellar models remain compatible with observational constraints from massive pulsars and gravitational wave events.

\subsection{The charge density relation}

To complement the equation of state describing the ultra-dense quark matter, an explicit prescription for the electric charge density $\rho_{\rm ch}$ is 
required when modelling charged stellar configurations. A commonly adopted ansatz, originally proposed in the context of general relativity by Ray 
\textit{et al.}~\cite{Ray:2003gt}, assumes that the charge distribution traces 
the local energy density. This proportionality is expressed as
\begin{equation}
    \rho_{\rm ch} = \beta\,\rho ,
    \label{charge-relation}
\end{equation}
where the dimensionless constant $\beta$ regulates the overall amount of electric charge contained in the fluid. The physical motivation for this choice is straightforward: regions
of higher mass-energy can support greater charge accumulation, making
this relation a natural first approximation for compact
stars~\cite{Arbanil:2013pua}.  In the ultra-dense
quark star interior specifically, this proportionality is further
motivated by the fact that the local quark number density — and hence
the net electric charge arising from the slight imbalance between
quark flavours — scales directly with the energy density at the
extreme conditions prevailing in the stellar core, making
$\rho_{\rm ch} = \beta\rho$ a physically transparent and
thermodynamically consistent first-order approximation for the charge
distribution~\cite{Ray:2003gt,Negreiros:2009fd}. 

This charge profile has been widely employed in recent investigations of charged quark stars within modified gravity frameworks, including analyses in 
metric $f(R)$ gravity~\cite{Pretel:2022rwx}, $f(R,T)$ gravity~\cite{Pretel:2022dbx}, 
and in the regularized $4D$ Einstein-Gauss-Bonnet theory \cite{Gammon:2024gij,Pretel:2021czp}. Its use in the present EMSG context facilitates a
consistent comparison with these previous studies while providing a simple, physically motivated mechanism for incorporating electric charge into the
stellar structure equations. We acknowledge that this proportionality is a phenomenological ansatz rather than a result derived from a first-principles QCD calculation. A rigorous microscopic determination of the charge distribution in dense quark matter remains an open problem in the field; the ansatz $\rho_{\rm ch} = \beta\rho$ represents the standard approach adopted across the charged compact star literature~\cite{Ray:2003gt,Negreiros:2009fd,Arbanil:2013pua,Pretel:2022rwx,Pretel:2022dbx,Gammon:2024gij}, and its use here is consistent with this established practice. Exploring more sophisticated charge distributions, potentially informed by QCD calculations at finite density, would be a valuable direction for future work. Throughout this work we fix the charge fraction at $\beta=0.5$, a representative intermediate value lying well within the physically admissible range $0\le\beta<1$ adopted in the charged compact star literature~\cite{Ray:2003gt,Arbanil:2013pua}, for which $\beta=0$ recovers the neutral limit and $\beta\to1$ approaches the extremal limit ($Q/M\to1$). This choice yields a sizeable yet sub-extremal electric charge ($Q/M\sim0.6$), large enough to exhibit the structural role of the Coulomb sector while keeping the configurations gravitationally bound. Increasing the charge fraction enhances the electric charge and the associated outward Coulomb support, raising the maximum mass and radius; this effect is quantified by the neutral ($\beta=0$) and charged ($\beta=0.5$) sequences in Tables~\ref{table1} and~\ref{table2}. For example, at EMSG coupling $\alpha=0$ and $\bar{\lambda}=0.1$ the maximum mass increases from $M_{\max}^{\rm n}=2.26\,M_\odot$ to $M_{\max}^{\rm ch}=2.88\,M_\odot$ when $\beta$ is raised from $0$ to $0.5$. A systematic scan over intermediate $\beta$ would provide a finer sensitivity map and is left for future work.

\begin{figure}[h]
    \centering
    \includegraphics[width = 8.7cm]{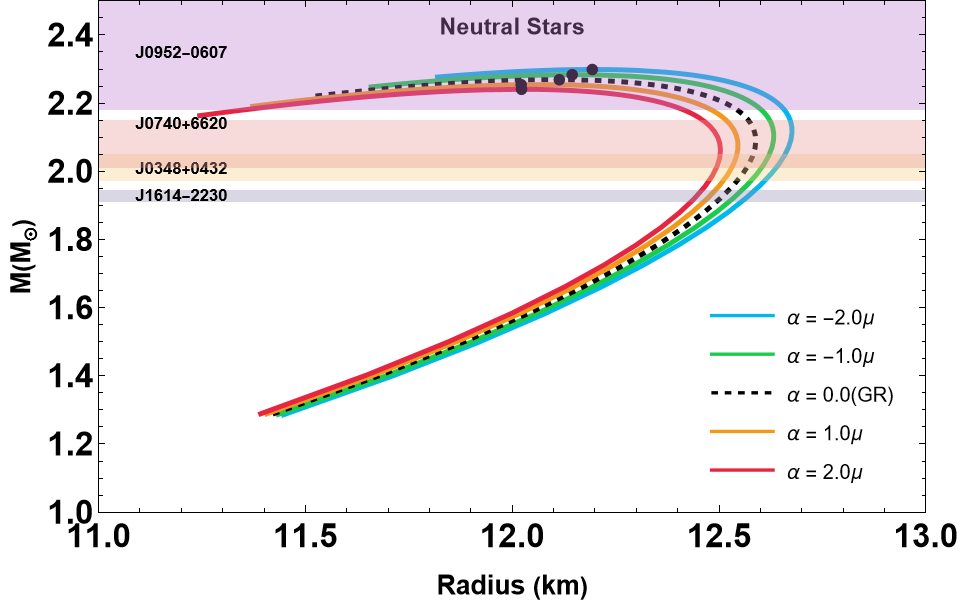}
    \includegraphics[width = 8.46cm]{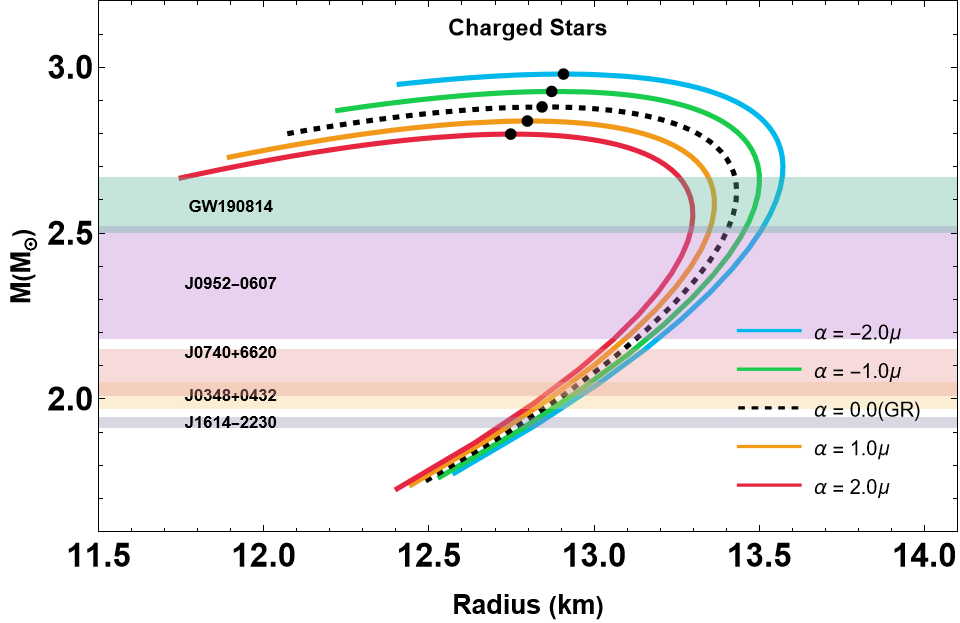}\\
    \includegraphics[width = 8.7cm]{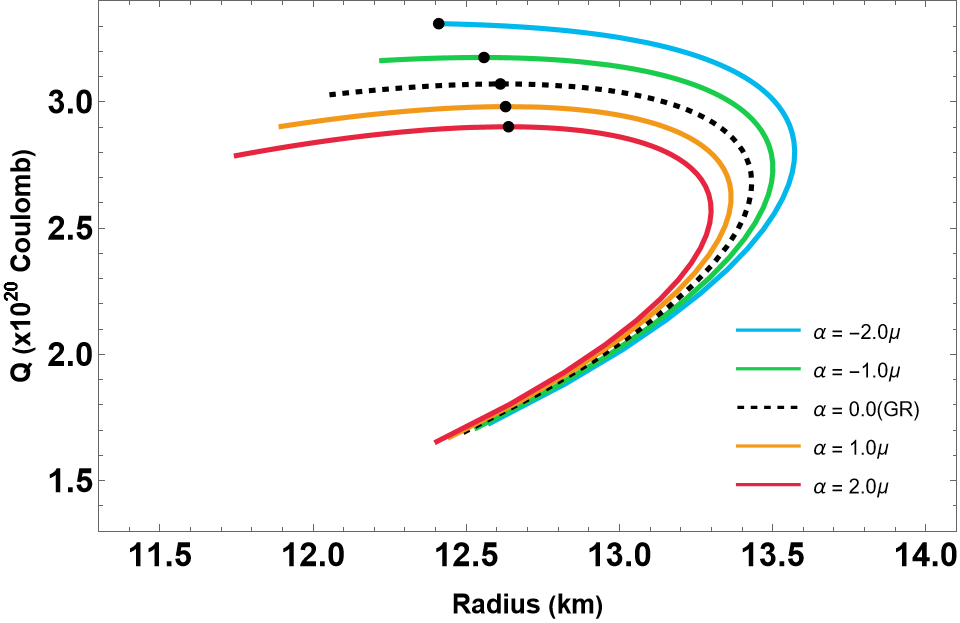}
    \includegraphics[width = 8.7cm]{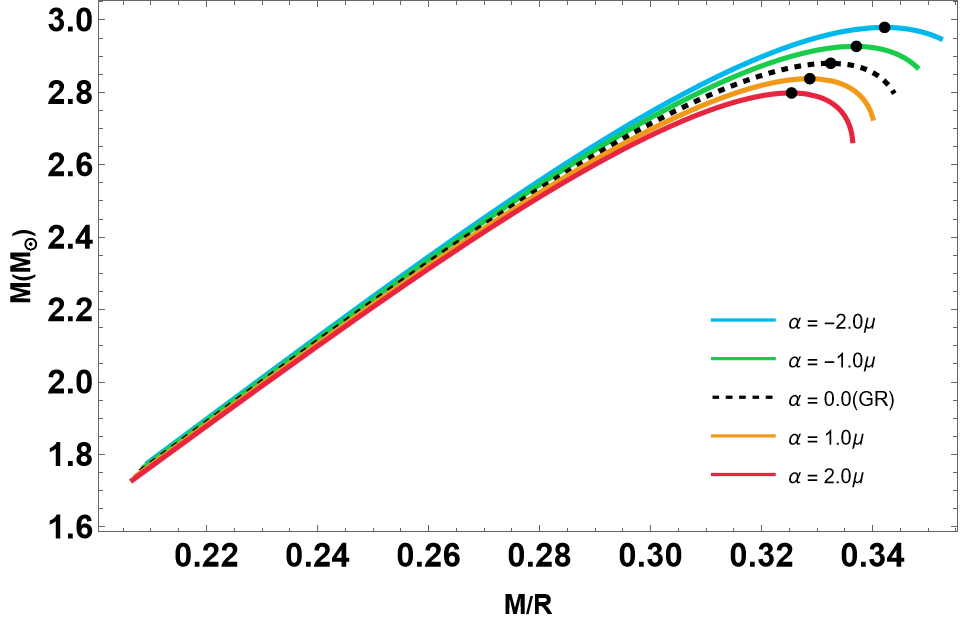}
    \caption[Mass-radius and charge-radius relations for varying EMSG coupling]{  Mass versus radius and charge versus radius relations for quark stars in energy-momentum squared gravity (EMSG), computed using the unified interacting quark matter equation of state. Each curve corresponds to a different value of the matter-geometry coupling parameter $\alpha$, expressed in units of $10^{-38}~\text{cm}^3~\text{erg}^{-1}$: $\alpha = -2$ (cyan), $-1$ (green), $0$ (black dashed, corresponding to general relativity), $+1$ (orange), and $+2$ (red). All sequences are generated with the interaction parameter $\bar{\lambda} = 0.1$, the charge fraction $\beta=0.5$ and bag constant $B_{\rm eff} = 60~\text{MeV}~\text{fm}^{-3}$. \textit{Top left}: Mass-radius curves for neutral stars. \textit{Top right}: Mass-radius curves for charged stars with a charge density proportional to the energy density. Horizontal shaded bands show observational constraints from PSR J1614-2230 with $1.97 \pm 0.04 \,M_{\odot}$ \cite{Ozel:2010bz},   PSR J0952$-$0607 with $M = 2.35 \pm 0.17\,M_{\odot}$ \cite{Romani:2022jhd}, PSR J0740+6620 with $M = 2.08^{+0.07}_{-0.07}\,M_{\odot}$ \cite{Fonseca:2021wxt}, 
    PSR J0348+0432 with $M = 2.01 \pm 0.04\,M_{\odot}$ \cite{Antoniadis:2013pzd}, and the mass range associated with the secondary component of} GW190814 \cite{LIGOScientific:2020zkf}. \textit{Bottom left}: Total electric charge as a function of radius. \textit{Bottom right}: Compactness $M/R$ as a function of gravitational mass. Black circles indicate the maximum mass point on each sequence for the corresponding parameter choice.  
    \label{fig1}
\end{figure}

\begin{figure}[h]
    \centering
    \includegraphics[width = 8.59cm]{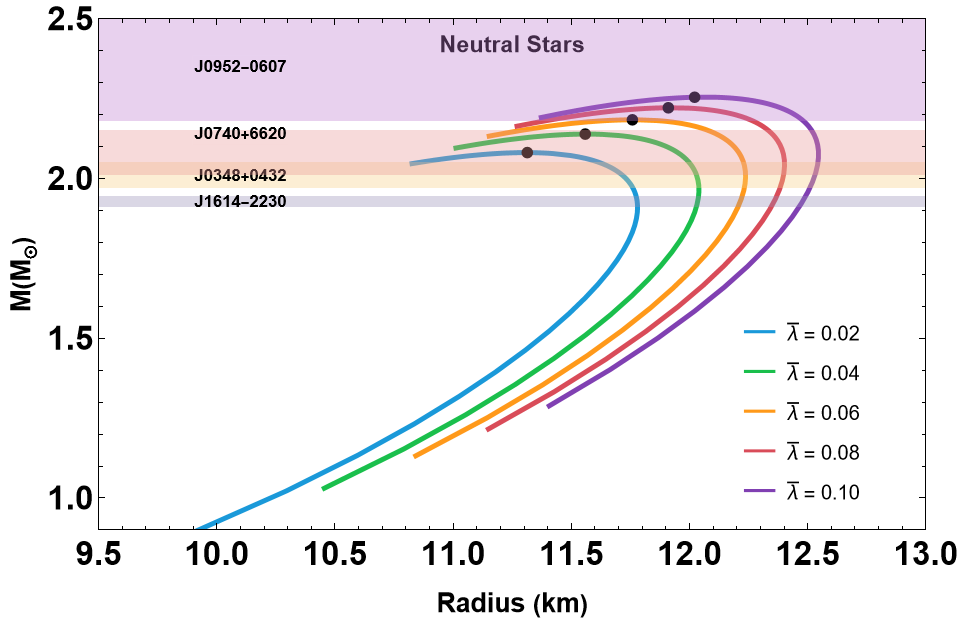}
    \includegraphics[width = 8.5cm]{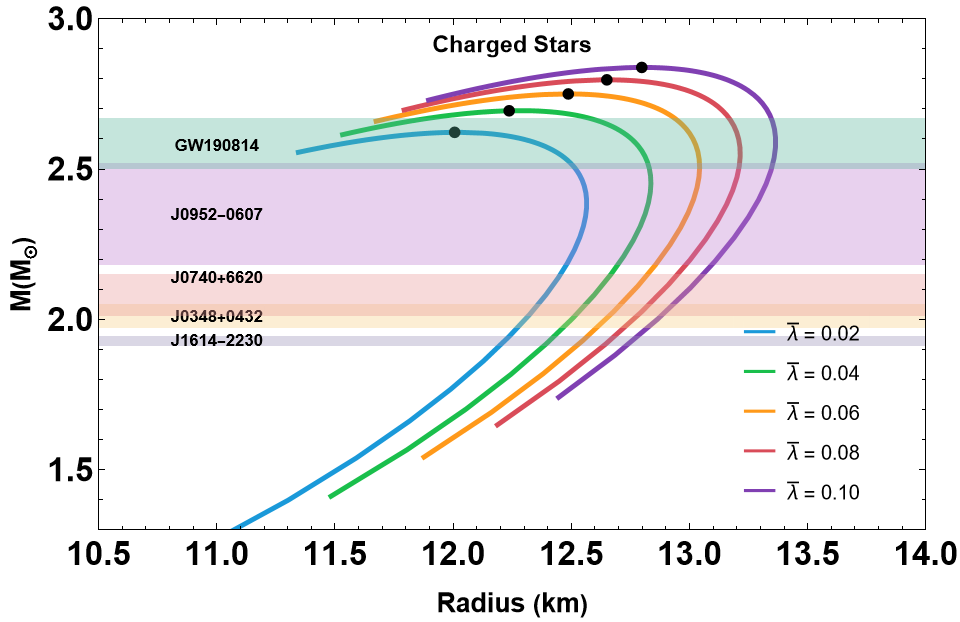}
    \includegraphics[width = 8.58cm]{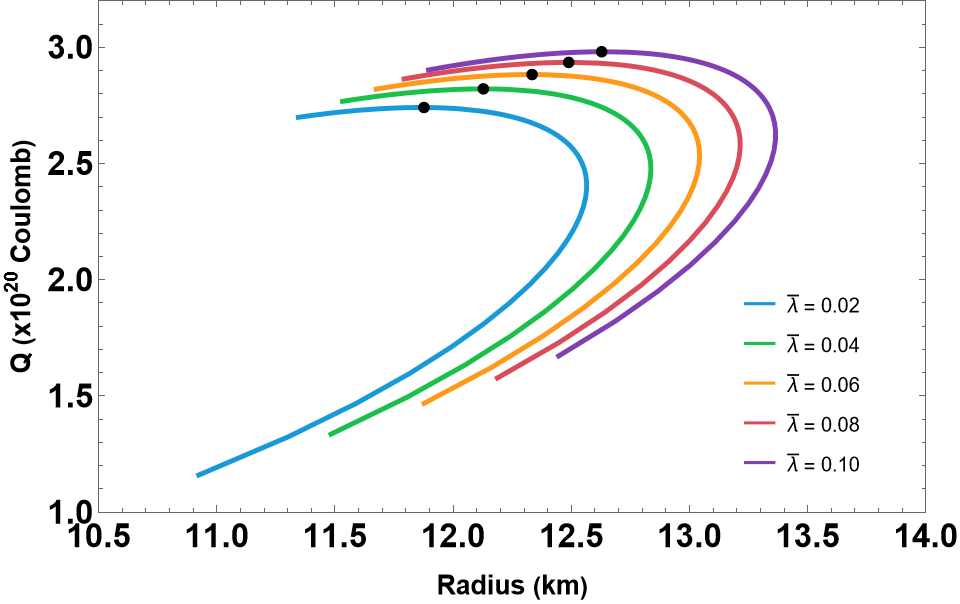}
    \includegraphics[width = 8.5cm]{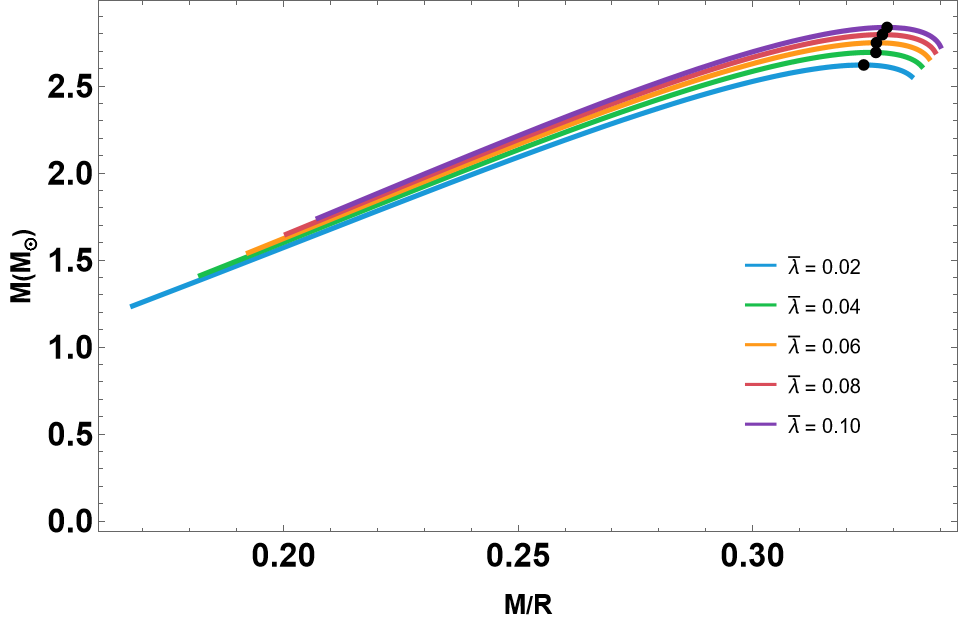}
    \caption[Mass-radius and charge-radius relations for varying interaction strength]{ Mass versus radius and charge versus radius relations for quark star configurations in energy-momentum squared gravity, obtained by varying the interaction parameter $\bar{\lambda}$ of the unified interacting quark matter equation of state. All models employ $B_{\text{eff}} = 60~\text{MeV}~\text{fm}^{-3}$, a fixed matter-geometry coupling $\alpha = 1.0\mu$ with $\mu = 10^{-38}~\text{cm}^3~\text{erg}^{-1}$, and the charge fraction $\beta=0.5$. Five representative values of $\bar{\lambda}$ are shown: $0.02$ (cyan), $0.04$ (green), $0.06$ (yellow), $0.08$ (orange), and $0.10$ (purple). \textit{Upper left panel}: Mass-radius sequences for electrically neutral stars. \textit{Upper right panel}: Mass-radius sequences for charged stars. Horizontal shaded bands indicate observational constraints from  PSR J1614-2230 \cite{Ozel:2010bz}, 
    J0348+0432 \cite{Antoniadis:2013pzd}, J0740+6620 \cite{Fonseca:2021wxt}, J0952-0607 \cite{Romani:2022jhd}, and the mass range associated with the secondary component of GW190814 \cite{LIGOScientific:2020zkf}. \textit{Lower left panel}: Total electric charge as a function of radius for the charged configurations. \textit{Lower right panel}: Compactness $M/R$ plotted against gravitational mass. Black circles mark the maximum mass configuration for each choice of $\bar{\lambda}$. }
    \label{fig2}
\end{figure}


 \section{Numerical results}\label{sec:numerical} 

In this section, we present the numerical solutions of the modified
TOV equations for charged quark stars within the EMSG framework.
Using the unified interacting quark matter equation of state together
with a density-proportional charge profile, we examine how the
coupling parameter $\alpha$, the interaction parameter $\bar{\lambda}$,
and the charge fraction $\beta$ influence the global stellar
properties. In particular, we analyze the resulting mass-radius
relations, total electric charge, compactness, and internal stability
indicators characterizing the equilibrium sequences obtained in this
model.  The numerical integrations are carried out over
the following parameter ranges: the EMSG coupling strength takes the
values $\alpha \in \{-2.0,\,-1.0,\,0.0,\,+1.0,\,+2.0\} \times
10^{-38}~\text{cm}^3~\text{erg}^{-1}$, the quark matter interaction
parameter spans $\bar{\lambda} \in \{0.02,\,0.04,\,0.06,\,0.08,\,
0.10\}$, and the charge fraction is fixed at $\beta = 0.5$ throughout,
with bag constant $B_{\rm eff} = 60~\text{MeV\,fm}^{-3}$. For every
combination of these parameters, we have verified that the central
energy density $\rho_c$ and the central pressure $p_c$ remain strictly
positive, confirming that all computed configurations correspond to
physically admissible stellar interiors with no pathological behaviour
at the origin.

\subsection{Influence of the EMSG coupling parameter $\alpha$ on the structure of quark stars}

 Varying the EMSG coupling constant $\alpha$ produces clear and
systematic changes in the equilibrium structure of both neutral
and charged quark stars, as illustrated in Fig.~\ref{fig1} and
summarized in Table~\ref{table1}.  Negative values of $\alpha$ strengthen the effective pressure contribution generated by the nonlinear matter-geometry coupling, enabling the star to 
support larger gravitational masses and radii at lower central densities. This trend is visible in the neutral sequences, but it becomes even more pronounced when electric charge is included. The Coulomb repulsion amplifies the stabilizing effect of negative $\alpha$, pushing the 
maximum mass of charged stars well beyond the general relativistic limit—for example, allowing configurations with $\alpha = -2\mu$ to attain nearly
$3M_\odot$ with compactness values approaching $M/R\simeq 0.34$,
a regime compatible with the mass range of the secondary compact
object in GW190814~\cite{LIGOScientific:2020zkf}, whose precise
nature remains uncertain but which has been widely discussed as a
possible quark star or heavy neutron star candidate in the
literature~\cite{Pretel:2023avv,Tangphati:2022acb,Dayanandan:2025vyw}. A black-hole interpretation of this object cannot be excluded either. Neutral stars over the same parameter range remain bounded by 
$M_{\rm max}\lesssim 2.3M_\odot$, demonstrating that while modifications to gravity alone enhance stability, electromagnetic forces supply an additional outward contribution required to achieve the highest observed stellar masses.

The influence of $\alpha$ is also evident in the total electric charge profiles. As shown in the bottom-left panel of Fig.~\ref{fig1}, the charge distribution increases monotonically with radius. It reaches its maximum at the stellar surface, consistent with the proportionality relation  $\rho_{\rm ch}=\beta\rho$. For negative $\alpha$, the enhanced pressure support allows the star to extend to slightly larger radii at comparable central density, resulting in higher total charge 
values—for instance, $Q_{\max}\approx 3.3\times10^{20}\,{\rm C}$ at $\alpha=-2\mu$. As $\alpha$ increases toward positive values, both the stellar radius and the integrated charge diminish systematically, reducing the surface charge to $Q_{\max}\approx 2.9\times10^{20}\,{\rm C}$ for $\alpha=+2\mu$. These variations capture the combined role of matter-geometry
coupling and the charge-density prescription in shaping the star's
electromagnetic content.  The surface charge values obtained here, $Q_{\max} \sim 2.9$--$3.3 \times 10^{20}$~C, are
consistent with those reported in recent investigations of charged
compact stars in modified gravity frameworks. Studies of charged
compact stars in non-minimally coupled gravity and in
matter-geometry coupled theories have reported surface charges of
comparable order of
magnitude~\cite{Sharif:2023vpb,Naseer:2024cto}, as have analyses
of charged quark stars in $f(R)$ and $f(R,T)$
gravity~\cite{Pretel:2022rwx,Pretel:2022dbx} and in regularized
four-dimensional Einstein-Gauss-Bonnet
theory~\cite{Pretel:2021czp}. The agreement across these independent frameworks indicates that the charge magnitudes obtained in the present work are comparable to those adopted in the broad class of charged compact-star models explored in the recent literature. We emphasize, however, that a net charge of this order is an idealization: a surface charge $Q\sim3\times10^{20}~{\rm C}$ on a star of radius $R\sim13~{\rm km}$ produces a surface field $E_{\rm surf}\sim1.7\times10^{22}~{\rm V\,m^{-1}}$, exceeding the Schwinger critical field $E_{\rm c}\simeq1.3\times10^{18}~{\rm V\,m^{-1}}$ for electron--positron pair production by roughly four orders of magnitude. In a realistic quantum vacuum such fields would drive prolific pair creation and rapid discharge, so charges of this magnitude are not expected to be sustained. The maximal charges adopted here should therefore be understood as bracketing the largest possible influence of electric charge on the equilibrium structure within EMSG, rather than as necessarily sustainable, globally unscreened charges in isolated compact stars. A realistic star may carry a much smaller net charge, and a quantitative assessment of how these trends persist would require a dedicated lower-charge or discharge-regulated analysis, beyond the scope of the present equilibrium treatment.

Across the full range of $\alpha$, all computed stellar sequences remain consistent with current 
astrophysical constraints, including the mass and radius measurements from NICER for  PSR~J1614$-$2230, J0348$+$0432, J0740$+$6620, and J0952$-$0607, as well as the bounds 
inferred from multimessenger gravitational-wave observations. Collectively, these results  highlight the complementary influence of modified gravity and moderate electric charge in 
governing the structure of ultra-dense matter, and emphasizing the potential of observational data to distinguish general relativity from alternative theories such as EMSG in the 
strong-field regime.

\subsection{Influence of the interaction parameter $\bar{\lambda}$ on the structure of quark stars}

The interaction parameter $\bar{\lambda}$ of the unified interacting quark matter equation of state 
plays a central role in determining the stiffness of dense matter and, consequently, the global 
properties of quark stars. The trends associated with varying $\bar{\lambda}$ are shown in Fig.~\ref{fig2} and summarized in Table~\ref{table2}. As shown in Table~\ref{table2}, the central pressure
$p_c$ at the maximum-mass configuration decreases from
$241.17~\text{MeV\,fm}^{-3}$ at $\bar{\lambda}=0.02$ to
$220.39~\text{MeV\,fm}^{-3}$ at $\bar{\lambda}=0.10$, reflecting
the progressive stiffening of the equation of state and the
associated reduction in central pressure required to support the
maximum-mass star as the interaction parameter increases. Smaller values of $\bar{\lambda}$ correspond to a softer equation of 
state, yielding stars that are less massive and more compact. In contrast, larger values lead to 
progressively stiffer matter capable of supporting higher gravitational masses and slightly 
expanded radii. This behavior is clearly visible in the neutral sequences: the maximum mass 
increases from $M_{\rm max}^n=2.08\,M_\odot$ at $\bar{\lambda}=0.02$ to $2.25\,M_\odot$ at 
$\bar{\lambda}=0.10$, accompanied by a systematic decrease in the central energy density at the 
stability limit.

The impact of $\bar{\lambda}$ becomes more pronounced for charged stars. As the parameter 
increases, the additional matter stiffness enhances the support provided by the Coulomb field, 
allowing charged configurations to reach larger masses within the same gravitational framework. 
For instance, the maximum mass rises from $M_{\rm max}^{\rm ch}=2.62\,M_\odot$ at $\bar{\lambda}=0.02$
to $2.84\,M_\odot$ at $\bar{\lambda}=0.10$, while the corresponding stellar radius expands from
approximately $12.0$ km to $12.8$ km. These variations reflect the joint influence of the quark
matter interaction strength and the electric charge contributions, both of which work to stiffen 
the stellar interior and suppress excessive compaction.

The total charge profiles shown in the lower-left panel of Fig.~2 further support this trend. 
Higher values of $\bar{\lambda}$ yield configurations that sustain larger integrated charge
and correspondingly higher surface charge, with $Q_{\max}$ increasing monotonically 
with $\bar{\lambda}$, from $2.74\times10^{20}~\text{C}$ at $\bar{\lambda}=0.02$ to 
$2.98\times10^{20}~\text{C}$ at $\bar{\lambda}=0.10$. This behavior parallels the 
changes observed in the mass-radius relations and underscores the sensitivity of the Coulomb 
sector to the stiffness of quark matter.

Importantly, all stellar sequences across the considered range of $\bar{\lambda}$ remain compatible 
with current observational constraints. Neutral configurations comfortably satisfy the mass 
measurements from NICER and radio pulsars, while the charged models with moderate 
$\bar{\lambda}$ values extend into the mass range associated with the secondary component, under a possible compact-star interpretation,
in GW190814. Taken together, these results highlight the role of quark matter interactions in 
shaping the macroscopic properties of quark stars and demonstrate how varying 
$\bar{\lambda}$ provides a complementary mechanism—alongside the EMSG coupling $\alpha$ and 
electric charge fraction $\beta$---for accessing the broader parameter space of ultra-dense 
compact stars in modified gravity.

\begin{figure}[h]
    \centering
    \includegraphics[width = 8.5cm]{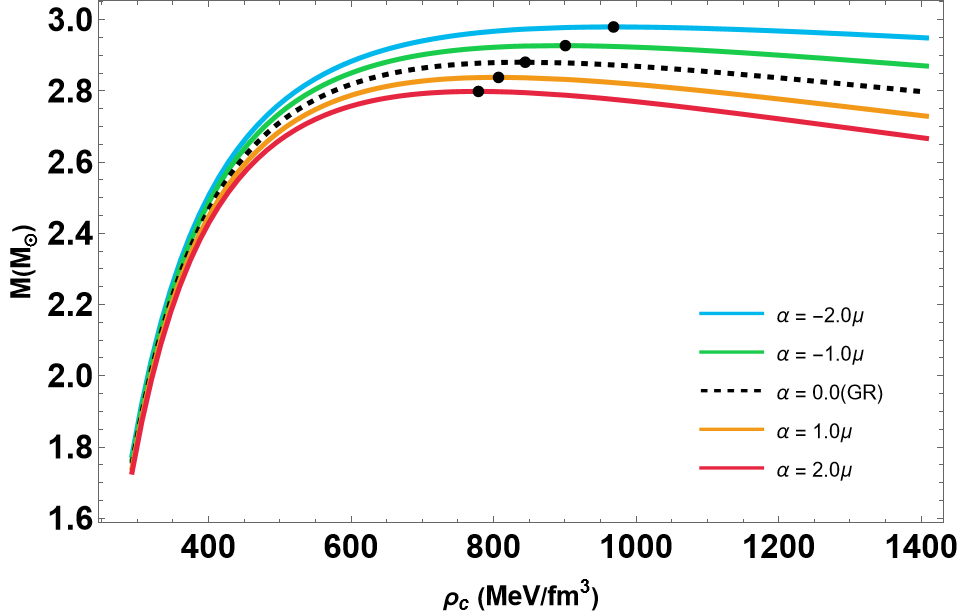}
    \includegraphics[width = 8.5cm]{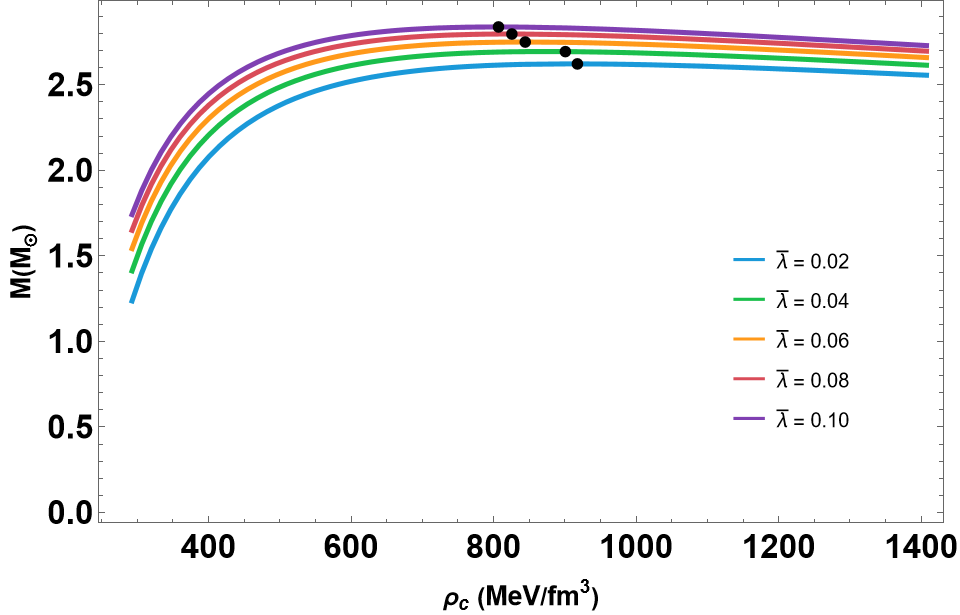}
    \caption[Mass-central-density curves for charged quark stars]{Gravitational mass $M$ as a function of the central energy density $\rho_c$ for charged quark star configurations in EMSG. The parameter spaces used in the left and right panels are identical to those adopted in Fig.~\ref{fig1} and Fig.~\ref{fig2}, respectively. \textbf{Left panel:} Mass--central-density curves for different values of the EMSG coupling parameter $\alpha$. \textbf{Right panel:} Corresponding sequences obtained by varying the quark-matter interaction parameter $\bar{\lambda}$. Black dots denote the maximum-mass configurations along each sequence.}
    \label{fig3}
\end{figure}

\begin{table}[t]
\centering
\caption[Maximum-mass properties for varying EMSG coupling]{ Key stellar properties at maximum mass for quark stars in energy-momentum squared gravity. Columns show the EMSG coupling parameter
$\alpha$ (in units of $10^{-38}~\text{cm}^3~\text{erg}^{-1}$), maximum masses for neutral ($M_{\max}^{\text{n}}$) and charged ($M_{\max}^{\text{ch}}$) stars, stellar radius ($R_{\max}^{\text{ch}}$), central density ($\rho_c$),  central pressure ($p_c$), maximum charge ($Q_{\max}$), and compactness $(M/R)_{\max}^{\text{ch}}$. All sequences computed with $\bar{\lambda} = 0.1$, charge fraction $\beta=0.5$ and $B_{\text{eff}} = 60~\text{MeV}~\text{fm}^{-3}$. }
\label{table1}
\renewcommand{\arraystretch}{1.3}
\begin{tabular}{@{}cccccccc@{}}
\toprule
$\alpha$
& $M_{\max}^{\text{n}}$
& $M_{\max}^{\text{ch}}$
& $R_{\max}^{\text{ch}}$
& $\rho_{c}$
& $p_{c}$
& $Q_{\max}$
& $(M/R)_{\max}^{\text{ch}}$ \\
$[10^{-38}~\text{cm}^3/\text{erg}]$
& $[M_{\odot}]$
& $[M_{\odot}]$
& $[\text{km}]$
& $[\text{MeV/fm}^{3}]$
& $[\text{MeV/fm}^{3}]$
& $[10^{20}~\text{C}]$
& \\
\midrule
$-2.0$  & 2.30 & 2.98 & 12.91 & 968 & 277.77 & 3.30 & 0.342 \\
$-1.0$  & 2.28 & 2.92 & 12.87 & 900 & 253.73 & 3.17 & 0.337 \\
$\phantom{-}0.0$  & 2.26 & 2.88 & 12.84 & 844 & 233.88 & 3.07 & 0.332 \\
$+1.0$  & 2.25 & 2.84 & 12.80 & 806 & 220.39 & 2.98 & 0.328 \\
$+2.0$  & 2.24 & 2.80 & 12.75 & 778 & 210.44 & 2.90 & 0.325 \\
\bottomrule
\end{tabular}
\end{table}

\begin{table}[t]
\centering
\caption[Maximum-mass properties for varying interaction strength]{Key stellar properties at maximum mass for quark star configurations in energy-momentum squared gravity as a function of the interaction parameter $\bar{\lambda}$. The table presents maximum masses for neutral ($M_{\max}^{\text{n}}$) and charged ($M_{\max}^{\text{ch}}$) configurations, along with the corresponding stellar radius ($R_{\max}^{\text{ch}}$), central energy density ($\rho_c$),  central pressure ($p_c$), total electric charge ($Q_{\max}$), and compactness $(M/R)_{\max}^{\text{ch}}$ at the maximum mass point. All sequences computed for $\alpha = 1.0\mu$ with $\mu = 10^{-38}~\text{cm}^3~\text{erg}^{-1}$, charge fraction $\beta = 0.5$, and bag constant $B_{\text{eff}} = 60~\text{MeV}~\text{fm}^{-3}$.}
\label{table2}
\renewcommand{\arraystretch}{1.3}
\begin{tabular}{@{}cccccccc@{}}
\toprule
$\bar{\lambda}$
& $M_{\max}^{\text{n}}$
& $M_{\max}^{\text{ch}}$
& $R_{\max}^{\text{ch}}$
& $\rho_{c}$
& $p_{c}$
& $Q_{\max}$
& $(M/R)_{\max}^{\text{ch}}$ \\
& $[M_{\odot}]$
& $[M_{\odot}]$
& $[\text{km}]$
& $[\text{MeV/fm}^{3}]$
& $[\text{MeV/fm}^{3}]$
& $[10^{20}~\text{C}]$
& \\
\midrule
$0.02$  & 2.08 & 2.62 & 12.00 & 917 & 241.17 & 2.74 & 0.324 \\
$0.04$  & 2.14 & 2.69 & 12.24 & 900 & 241.58 & 2.82 & 0.326 \\
$0.06$  & 2.18 & 2.75 & 12.49 & 844 & 226.73 & 2.88 & 0.326 \\
$0.08$  & 2.22 & 2.80 & 12.65 & 825 & 223.84 & 2.93 & 0.327 \\
$0.10$  & 2.25 & 2.84 & 12.80 & 806 & 220.39 & 2.98 & 0.328 \\
\bottomrule
\end{tabular}
\end{table}
To facilitate independent reproduction of the models presented here,
we highlight two representative maximum-mass configurations from
Table~\ref{table1}. For the strongest negative EMSG coupling
$\alpha = -2.0\,\mu$ with $\bar{\lambda} = 0.1$, $\beta = 0.5$,
and $B_{\rm eff} = 60~\text{MeV\,fm}^{-3}$, the maximum-mass
charged star has central density $\rho_c = 968~\text{MeV\,fm}^{-3}$,
central pressure $p_c = 277.77~\text{MeV\,fm}^{-3}$, gravitational
mass $M = 2.98\,M_{\odot}$, radius $R = 12.91~\text{km}$, and
total surface charge $Q = 3.30\times10^{20}~\text{C}$. The
corresponding general relativistic configuration ($\alpha = 0$)
yields $\rho_c = 844~\text{MeV\,fm}^{-3}$,
$p_c = 233.88~\text{MeV\,fm}^{-3}$, $M = 2.88\,M_{\odot}$,
$R = 12.84~\text{km}$, and $Q = 3.07\times10^{20}~\text{C}$.
These two models bracket the range of EMSG effects studied in
this work and provide sufficient information for independent
numerical verification.

\subsection{Radial charge density profile}

\begin{figure}[h]
    \centering
    \includegraphics[width=8.59cm]{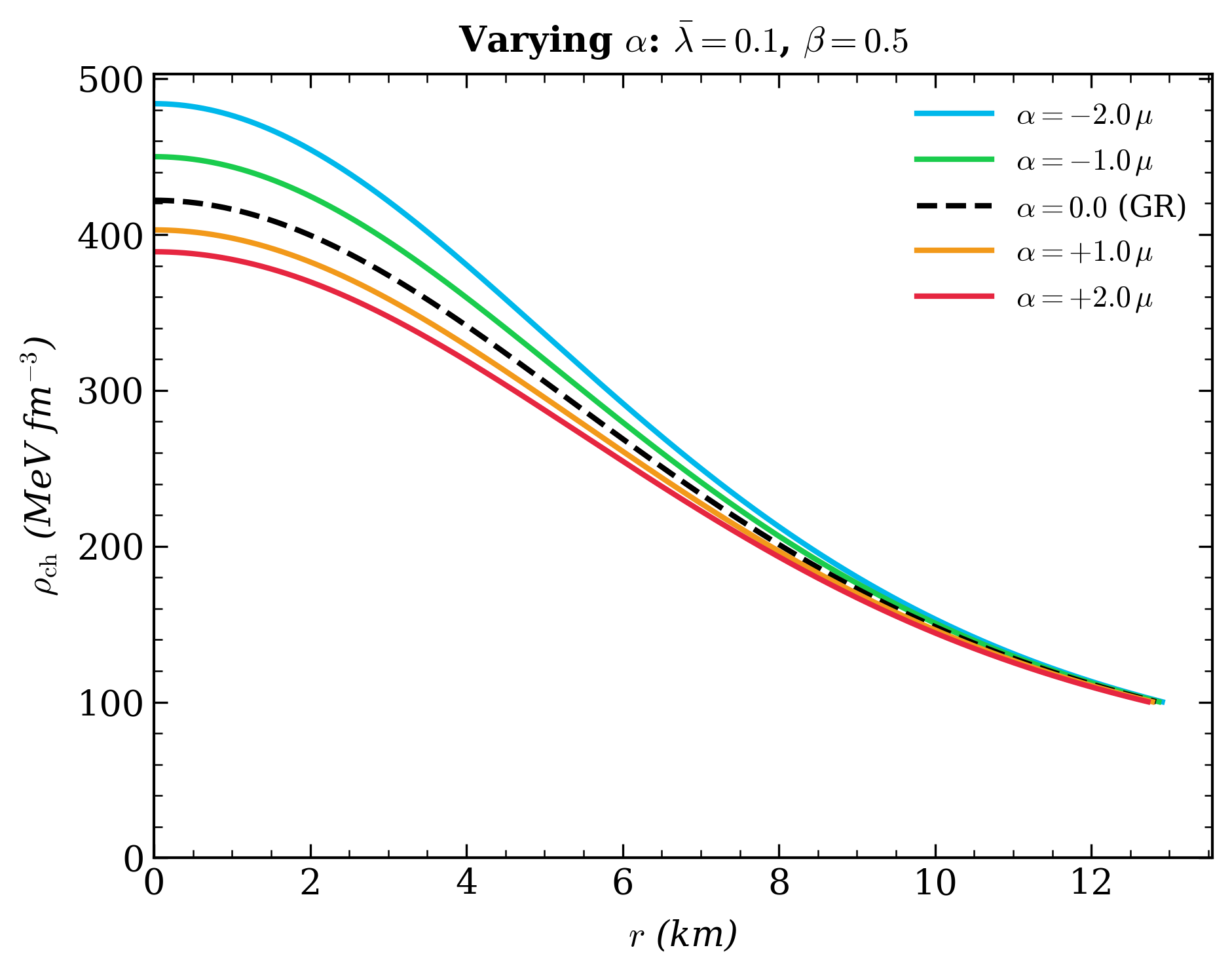}
    \includegraphics[width=8.50cm]{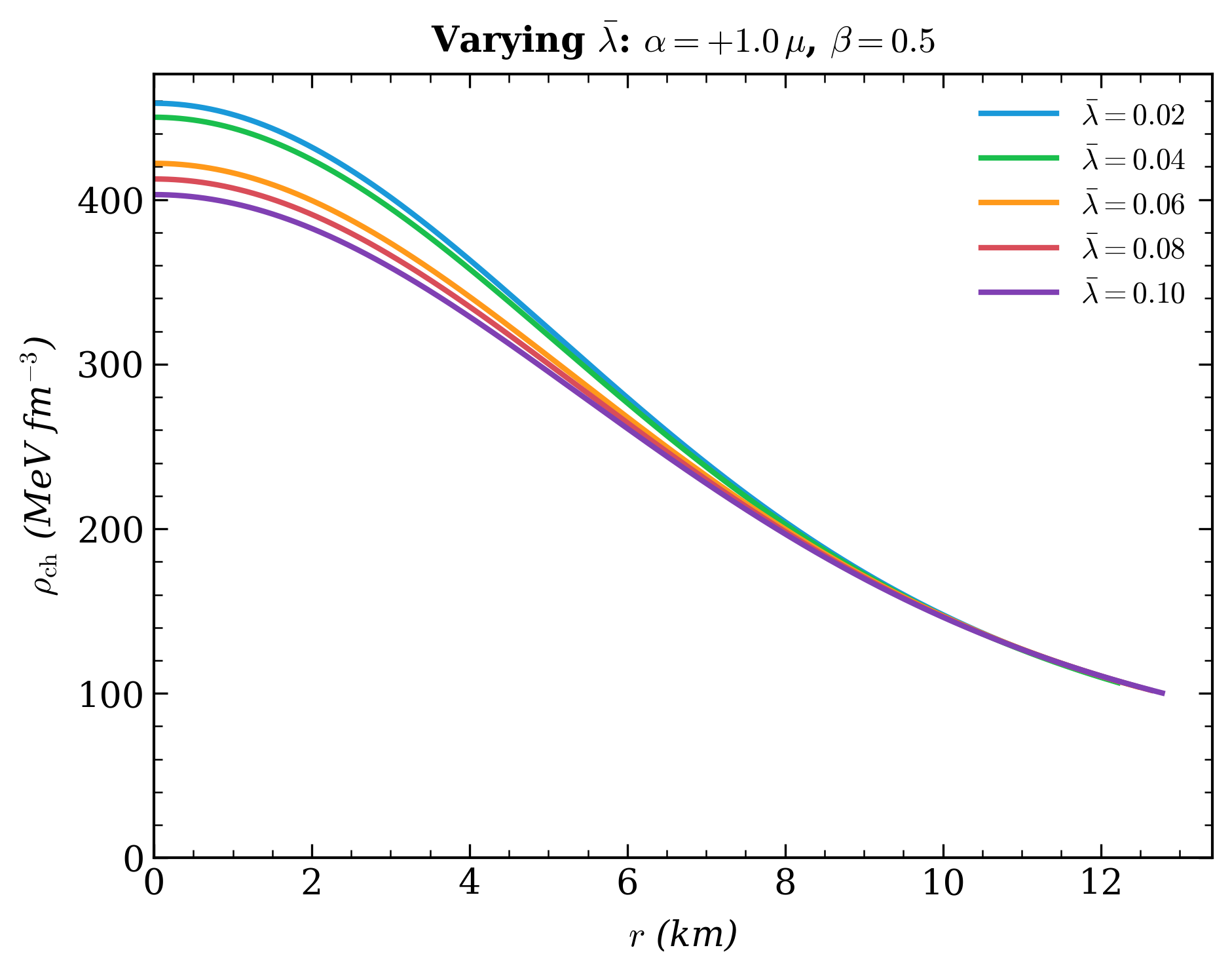}
    \caption[Radial profiles of the charge density]{ Radial profiles of the charge density
    $\rho_{\rm ch}(r) = \beta\,\rho(r)$ for charged quark star
    configurations in EMSG, each evaluated at the central density
    of the maximum-mass star for the corresponding parameter set.
    \textit{Left panel:} Dependence of $\rho_{\rm ch}(r)$ on the
    EMSG coupling parameter $\alpha$, with central densities taken
    from Table~\ref{table1} ($\bar{\lambda}=0.1$, $\beta=0.5$,
    $B_{\rm eff}=60~\text{MeV\,fm}^{-3}$).
    \textit{Right panel:} Corresponding profiles for varying
    interaction parameter $\bar{\lambda}$, with central densities
    from Table~\ref{table2} ($\alpha=+1.0\,\mu$, $\beta=0.5$).
    In all cases, $\rho_{\rm ch}$ is finite and smooth at the
    stellar centre, remains strictly positive throughout the
    interior, and decreases monotonically to a nonzero surface
    value characteristic of quark matter with a finite bag
    constant.}
    \label{fig4}
\end{figure}

Figure~\ref{fig4} shows the radial profile of the charge density
$\rho_{\rm ch}(r) = \beta\,\rho(r)$ obtained from the numerical
integration of the modified TOV equations, with each curve evaluated
at the central density corresponding to the maximum-mass configuration
for that parameter set, as listed in Tables~\ref{table1}
and~\ref{table2}. In both panels, $\rho_{\rm ch}$ attains its peak
value at the stellar centre, $\rho_{\rm ch}(0) = \beta\,\rho_c$,
where $\rho_c$ differs for each curve according to the maximum-mass
entry in the respective table, and decreases monotonically outward,
remaining strictly positive throughout the interior. The profile is
finite and smooth at $r = 0$, consistent with the regularity
condition imposed by the boundary condition $q(0) = 0$. At the
stellar surface, $\rho_{\rm ch}$ does not vanish but settles to a
nonzero value determined by the finite surface density of quark
matter, a characteristic feature of the bag-model equation of state
where pressure vanishes at a nonzero energy density set by
$B_{\rm eff}$. In the left panel, stronger negative values of
$\alpha$ correspond to higher maximum-mass central densities and
therefore produce larger central charge densities, while also
extending the star to slightly larger radii. In the right panel,
smaller values of $\bar{\lambda}$ are associated with softer equations
of state and higher maximum-mass central densities, yielding larger
central charge densities but shorter stellar radii. In all cases the
charge density remains positive definite, confirming that the adopted
ansatz $\rho_{\rm ch} = \beta\rho$ with $\beta > 0$ introduces a
physically consistent, positive-definite charge distribution
throughout the stellar interior.

\section{THE STATIC STABILITY CRITERION, ADIABATIC INDEX, AND SOUND VELOCITY}\label{sec:stability}

\subsection{Static stability from the $M\!-\!\rho_{c}$ relation}

 The equilibrium stability of charged quark stars in EMSG is intimately linked to the topology of the mass-central density relation, as shown in Figure~\ref{fig3}. For each parameter set, the locus of maximum mass marks the onset of radial instability, conforming to the classical stability argument developed by Harrison \textit{et al.}~\cite{1965gtgc.book.....H} and further established by Zeldovich and Novikov~\cite{1971reas.book.....Z}. When tracing the $M(\rho_c)$ curve upward, models on the rising branch ($dM/d\rho_c > 0$) are dynamically stable, while the descending branch ($dM/d\rho_c < 0$) corresponds to unstable configurations.

The effect of varying the EMSG coupling parameter $\alpha$ and the matter interaction strength $\bar{\lambda}$ is to shift the location and extent of the stable sequence systematically. Strongly negative $\alpha$ or enhanced interaction parameters yield higher maximum masses and expand the stable region toward lower central densities; conversely, positive $\alpha$ or softer interaction strengths curtail stability. This behavior reflects both the influence of nonlinear gravity and the underlying quark matter microphysics, reaffirming the utility of the turning-point criterion as indicated in Refs.~\cite{1965gtgc.book.....H,1971reas.book.....Z}  for diagnosing stellar stability in highly compact, strongly interacting regimes.

\begin{figure}[h]
    \centering
    \includegraphics[width = 8.59cm]{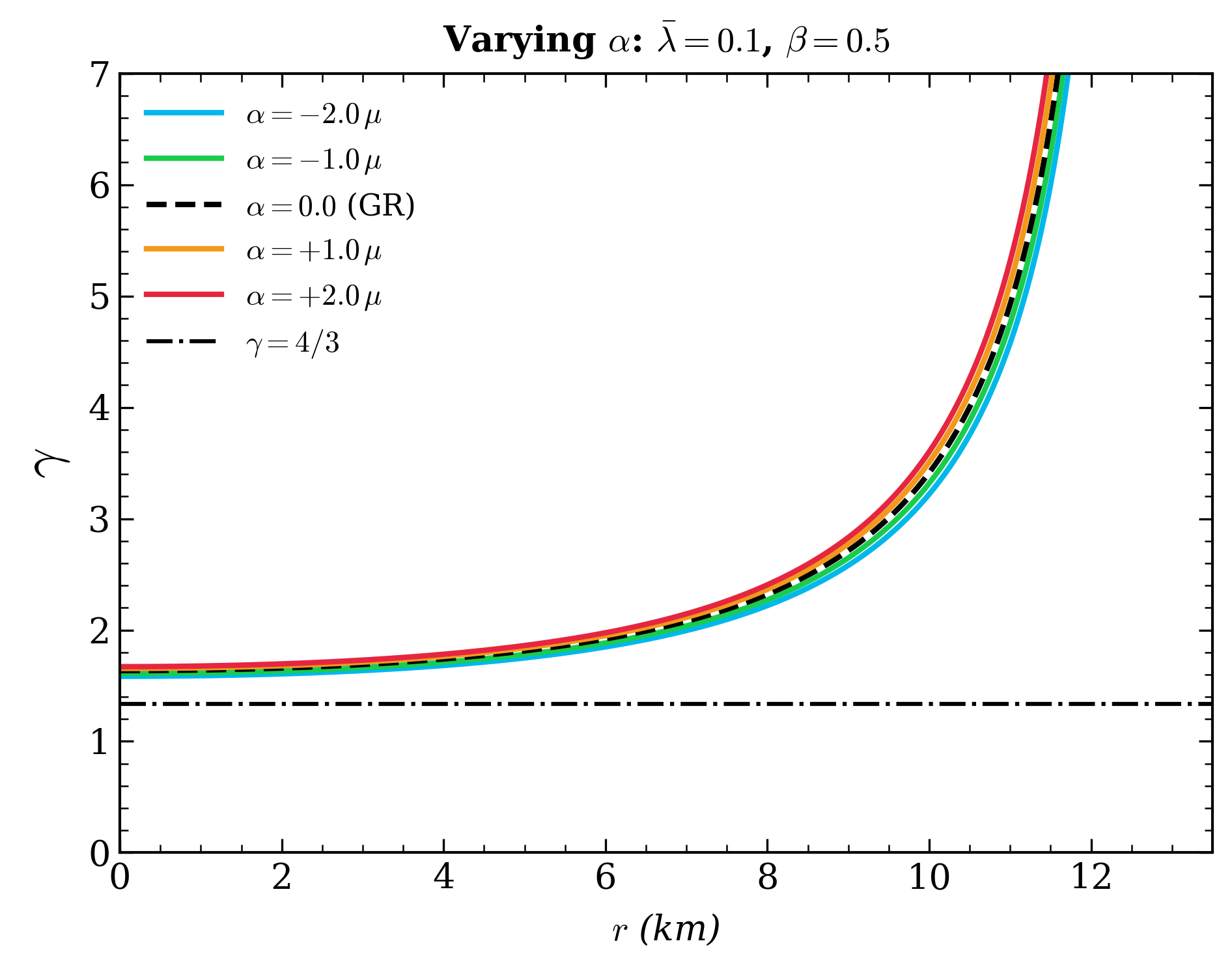}
    \includegraphics[width = 8.5cm]{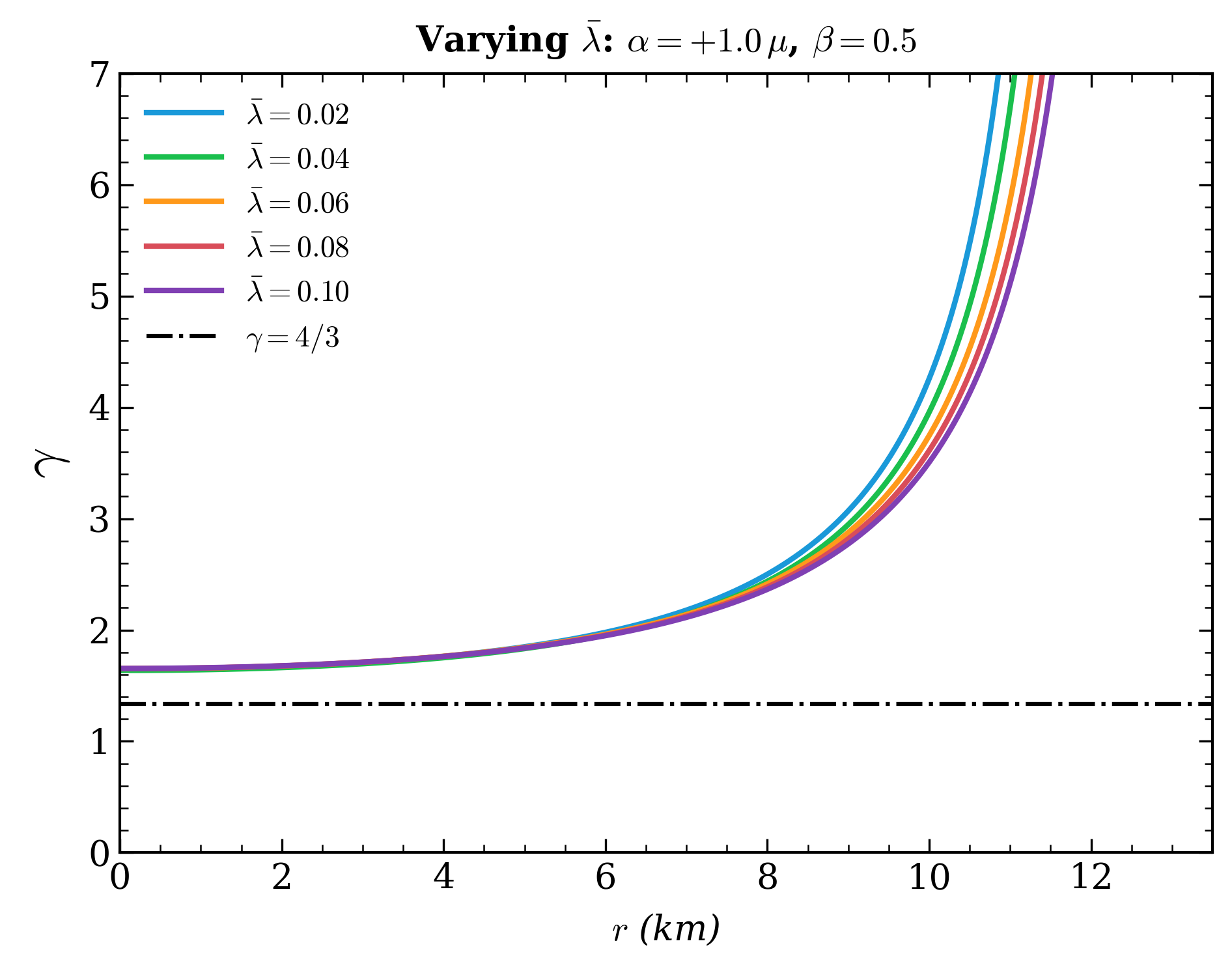}
    \caption[Radial profiles of the adiabatic index]{ Radial profiles of the adiabatic index $\gamma(r)$ for charged quark star configurations in energy-momentum squared gravity. The left panel shows how $\gamma$ varies as a function of radius for representative values of the EMSG coupling parameter $\alpha$, based on the parameter space adopted in Fig.~\ref{fig1}. In the right panel, the adiabatic index is plotted as a function of the quark matter interaction parameter $\bar{\lambda}$, with parameter choices matching those in Fig.~\ref{fig2}. In all cases, $\gamma$ rises steeply near the stellar surface and consistently remains above the relativistic instability limit ($\gamma = 4/3$, dash-dotted line), so that the explored models satisfy this necessary condition for stability against radial perturbations even in regimes of strong gravity and quark interactions.}
    \label{fig5}
\end{figure}

\begin{figure}[h]
    \centering
    \includegraphics[width = 8.58cm]{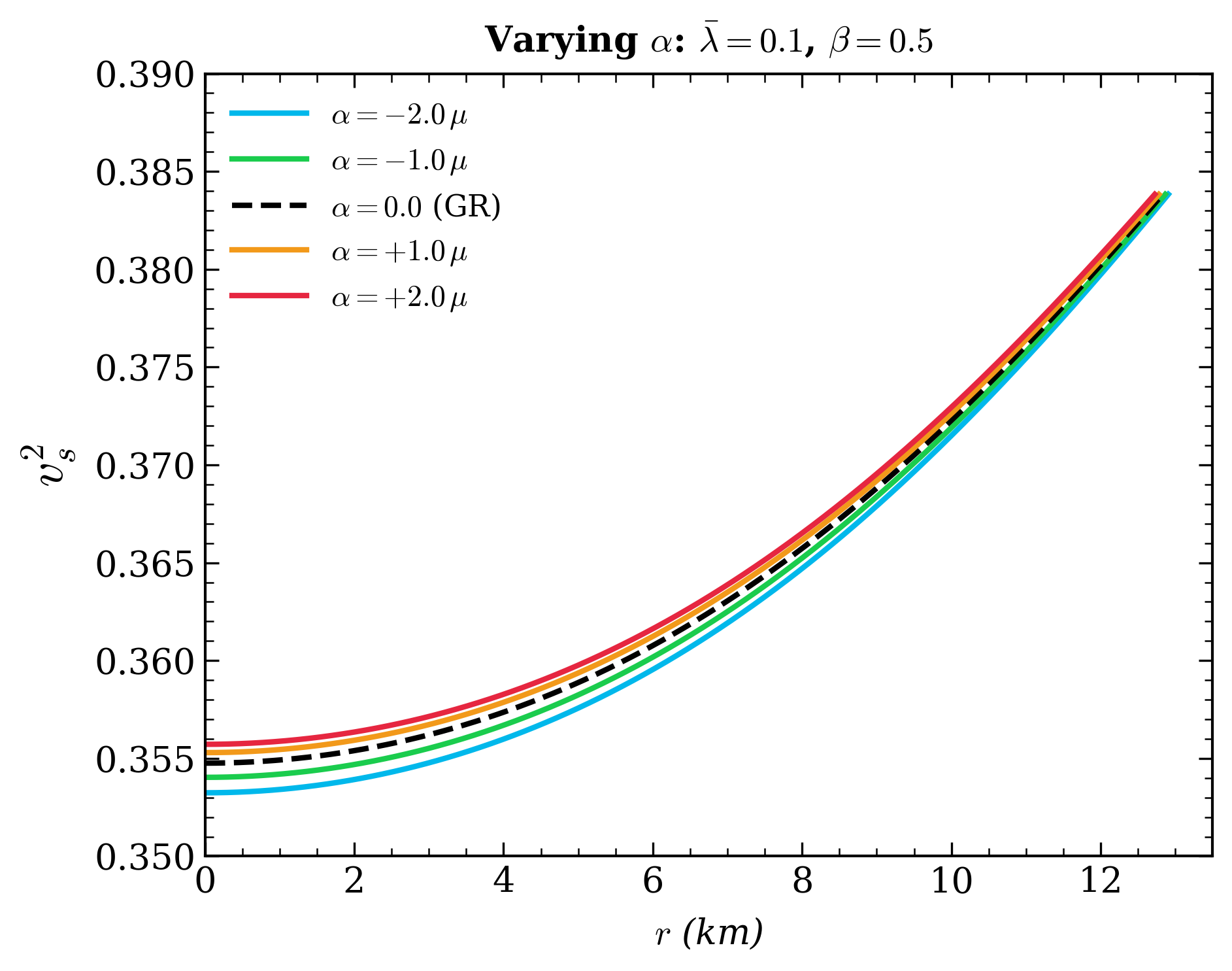}
    \includegraphics[width = 8.5cm]{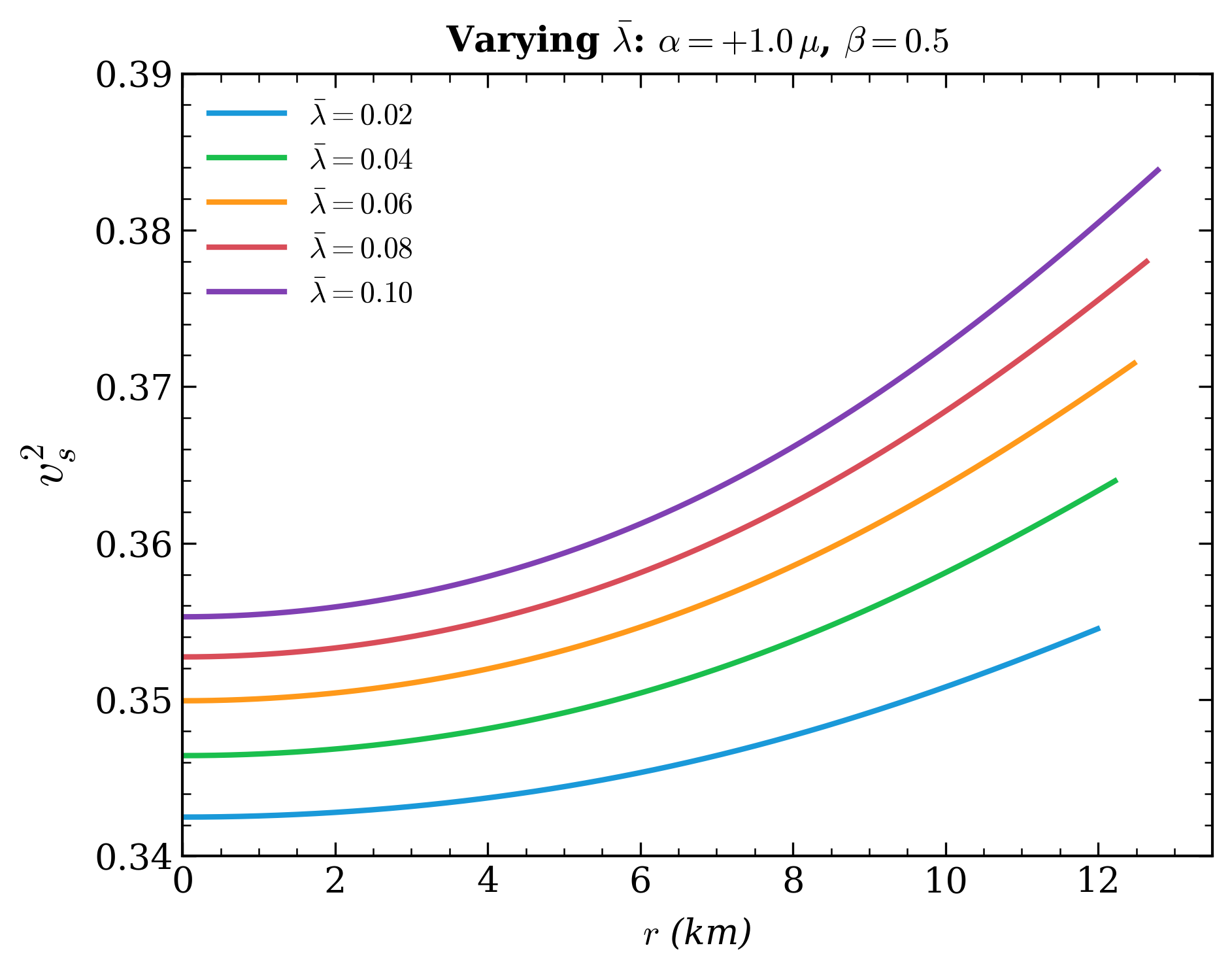}
    \caption[Radial profiles of the squared sound speed]{Radial profiles of the squared sound speed $v_s^{2} = dp/d\rho$ for charged quark star configurations in EMSG. \textit{Left panel:} Dependence of $v_s^{2}(r)$ on different values of the matter-geometry coupling parameter $\alpha$, using the same parameter space as in Fig.~\ref{fig1}. \textit{Right panel:} Corresponding sound-speed profiles for varying interaction strength $\bar{\lambda}$, with parameter choices matching those of Fig.~\ref{fig2}.
In all cases, $v_s^{2}$ increases gradually from the stellar core toward the surface and remains well below the causal limit $v_s^{2} = 1$, confirming that the explored models satisfy the causality condition throughout the interior.}
    \label{fig6}
\end{figure}

\subsection{Adiabatic index and dynamical stability}

The adiabatic index,
\begin{equation}
\gamma = \left(1 + \frac{\rho}{p}\right)\left(\frac{dp}{d\rho}\right) = \frac{\rho + p}{p}\left(\frac{dp}{d\rho}\right),
\end{equation}
plays a fundamental role in diagnosing the dynamical stability of relativistic stars against small radial perturbations. 
According to the classical analysis by Chandrasekhar~\cite{Chandrasekhar:1964zz}, a stellar configuration becomes dynamically unstable when the effective stiffness of matter falls below the critical threshold $\gamma < 4/3$, a limit characteristic of relativistically supported compact objects. Later refinements have emphasized that the precise stability boundary depends sensitively on the underlying microphysical equation of state and the strong-gravity environment~\cite{Moustakidis:2016ndw}. 
In modified gravity theories where energy-momentum conservation can be violated, additional corrections can also influence the radial stability condition~\cite{Maulana:2019sgd}.

Figure~\ref{fig5} presents the radial behaviour of the adiabatic index for charged quark star configurations in the EMSG framework, obtained using the same parameter sets employed in Figs.~1 and~2. 
In all cases, the profiles exhibit a monotonic rise from the core toward the stellar surface, consistently remaining above the relativistic instability threshold $\gamma = 4/3$. 
This trend reflects the progressive stiffening of quark matter at lower densities and indicates that the equilibrium sequences satisfy the necessary condition $\gamma > 4/3$ for stability against radial perturbations.

Variations in the EMSG coupling parameter $\alpha$ or in the quark-matter interaction strength $\bar{\lambda}$ primarily affect the outer layers of the star, with more negative $\alpha$ or larger $\bar{\lambda}$ producing slightly higher values of $\gamma$ near the surface. 
Nevertheless, no parameter choice within the explored domain yields regions with $\gamma < 4/3$, so that the necessary condition $\gamma > 4/3$ for radial stability is satisfied throughout their interiors.
This behaviour is entirely consistent with earlier analyses of quark stars in the EMSG framework~\cite{Banerjee:2025ewc,Dayanandan:2025vyw}. It demonstrates that the nonlinear matter-geometry coupling does not
introduce destabilizing effects for the parameter ranges considered
here.  We emphasize that all stability diagnostics
presented here --- the adiabatic index, the sound speed, and the
turning-point criterion --- are evaluated for the total effective
fluid that incorporates both the quark matter contributions and the
EMSG gravity correction terms, following the approach adopted in
Refs.~\cite{Naseer:2024jjy,2025PDU....5002133N}.

 \subsection{Sound speed and the causality condition}

The squared sound speed,
\begin{equation}
v_s^{2} = \frac{dp}{d\rho},
\end{equation}
provides a direct measure of the stiffness of the equation of state and must satisfy the causality requirement $v_s^{2} \leq 1$ throughout the stellar interior. Figure~\ref{fig6} illustrates the radial behaviour of $v_s^{2}$ for the charged quark star models considered in this work.

Across all parameter sets, the sound speed increases slightly from the stellar core toward the outer layers and remains consistently below the causal limit. Variations in the EMSG coupling parameter $\alpha$ (left panel) induce only minimal changes in the profiles, indicating that nonlinear matter-geometry effects do not significantly modify the microphysical stiffness of matter. Changes in the interaction parameter $\bar{\lambda}$ (right panel) generate a slightly wider spread in the curves, with larger $\bar{\lambda}$ corresponding to a stiffer response, yet still safely within the subluminal regime.

These results demonstrate that all configurations
examined---regardless of $\alpha$, $\bar{\lambda}$, or the presence
of electric charge---satisfy the causality condition, confirming the
physical viability of the charged quark star solutions obtained
within the EMSG framework.  We stress that the
stability assessment presented in this work is based exclusively on
static criteria: the turning-point condition applied to the
$M(\rho_c)$ relation, the adiabatic index condition $\gamma > 4/3$,
and the causality requirement $v_s^2 \leq 1$. A rigorous determination of dynamical stability would require a full radial oscillation (eigenfrequency) analysis --- solving the radial pulsation equations for the normal-mode spectrum $\omega_n^2$ --- that consistently incorporates the EMSG gravity corrections, the electromagnetic field, and the quark matter interactions. Such an analysis represents a technically demanding extension that lies outside the scope of the present study and is left for future work~\cite{Maulana:2019sgd}.

\section{Concluding Remarks}\label{sec:conclusion}

In this work, we have carried out a detailed investigation of electrically charged quark stars within the framework of energy-momentum squared gravity (EMSG), employing a unified interacting quark-matter equation of state that incorporates perturbative QCD effects, color superconductivity, and finite strange-quark-mass contributions. By extending the Maxwell-EMSG field equations to include nonlinear matter-geometry couplings and a density-dependent charge profile, we derived the corresponding stellar-structure equations and systematically analyzed the model-parameter dependence of equilibrium configurations. Our study demonstrates that the combined effects of electric charge and EMSG corrections can significantly influence the global properties of ultra-dense compact stars.

The mass-radius relations reveal that negative values of the EMSG coupling parameter $\alpha$ considerably enhance the pressure support generated by the quadratic matter term, enabling both neutral and charged stars to sustain larger gravitational masses at lower central densities. When a moderate electric charge is included, this stabilizing effect becomes even more pronounced, allowing maximum masses approaching $3M_{\odot}$ with compactness values near $M/R \simeq 0.34$. These results place charged quark stars in EMSG within the observationally relevant regime of heavy pulsars such as PSR J0952-0607 and, under a possible compact-star interpretation, the mass range associated with the secondary component of GW190814. Variations in the interaction parameter $\bar{\lambda}$ further broaden the accessible mass range, confirming that quark matter microphysics and nonlinear gravitational effects jointly determine the structure of dense stellar objects.

We also examined the stability of the resulting configurations using several complementary diagnostics. The turning-point criterion applied to the $M(\rho_{c})$ relation indicates that the onset of instability occurs at the maximum-mass point for each sequence, consistent with the classical analyses of Chandrasekhar and subsequent refinements. Moreover, the radial profiles of the adiabatic index $\gamma(r)$ remain above the relativistic instability threshold $\gamma = 4/3$ throughout the interior, regardless of the choice of $\alpha$ or $\bar{\lambda}$. This behaviour is further supported by the sound-speed analysis, which shows that $v_{s}^{2}$ always stays comfortably below the causal limit across the entire stellar radius. These complementary indicators --- the turning-point criterion, $\gamma > 4/3$, and $v_s^2 \le 1$ --- are necessary conditions for stability and are satisfied throughout the explored parameter space, indicating physically viable charged quark star configurations within the EMSG framework; a rigorous assessment of dynamical stability through a full radial oscillation (eigenfrequency) analysis is left for future work.

The broader implications of our results highlight the importance of nonlinear matter-geometry couplings in shaping the behaviour of strongly interacting systems. The ability of EMSG to support high-mass compact stars without violating causality or stability constraints provides a compelling extension to general relativity in the strong-field regime. At the same time, the observed sensitivity of stellar properties to the interaction strength $\bar{\lambda}$ underscores the necessity of consistent, phenomenologically motivated quark matter models when interpreting astrophysical observations. The unified interacting EoS employed here provides a versatile framework for studying matter under extreme conditions while remaining consistent with current multimessenger constraints.

Several avenues for future research naturally emerge from the present investigation. A detailed analysis of tidal deformability, particularly in the context of binary mergers observed by LIGO–Virgo–KAGRA, would help constrain the viable parameter ranges for $\alpha$, $\beta$, and $\bar{\lambda}$. Extensions to slowly- or rapidly-rotating configurations, universal relations for moment-of-inertia compactness, and the potential role of anisotropic pressures would also provide deeper insight into the phenomenology of quark stars in EMSG. Additionally, alternative charge distributions or more general matter Lagrangians could further clarify the interplay between nonlinear gravitational effects and electromagnetic contributions. These efforts would help determine whether EMSG leaves distinctive observational signatures that could distinguish it from general relativity using current and upcoming astrophysical measurements.

\section*{Acknowledgments}

The authors sincerely thank the anonymous reviewers for their valuable comments and constructive suggestions, which have significantly improved the manuscript's quality and clarity. J.R. acknowledges the Grants No. U2541210 of the National Natural Science Foundation of China (NSFC) and No. F-FA-2021-510 of the Uzbekistan Ministry of Innovative Development.

\bibliography{References}

\end{document}